\input amstex
\documentstyle{amsppt}
\nologo \NoRunningHeads \TagsOnRight\pagewidth{37pc}\pageheight{54pc}
$\qquad\qquad\qquad\qquad\qquad\qquad\qquad\qquad\qquad\qquad\qquad\qquad\qquad\qquad\qquad\qquad\qquad\text{UWThPh-2007-8}$
\bigskip\bigskip\bigskip
\topmatter
\title
Functional integration and gauge ambiguities in generalized abelian gauge theories
\endtitle
\author
Gerald Kelnhofer
\endauthor
\affil Faculty of Physics\\ University of Vienna\\Boltzmanngasse 5, A-1090 Vienna\\ Austria
\endaffil

\abstract We consider the covariant quantization of generalized abelian gauge theories on a closed and compact $n$-dimensional manifold whose
space of gauge invariant fields is the abelian group of Cheeger-Simons differential characters. The space of gauge fields is shown to be a
non-trivial bundle over the orbits of the subgroup of smooth Cheeger-Simons differential characters. Furthermore each orbit itself has the
structure of a bundle over a multi-dimensional torus. As a consequence there is a topological obstruction to the existence of a global gauge
fixing condition. A functional integral measure is proposed on the space of gauge fields which takes this problem into account and provides a
regularization of the gauge degrees of freedom. For the generalized $p$-form Maxwell theory closed expressions for all physical observables are
obtained. The Green\rq s functions are shown to be affected by the non-trivial bundle structure. Finally the vacuum expectation values of
circle-valued homomorphisms, including the Wilson operator for singular $p$-cycles of the manifold, are computed and selection rules are
derived.
\endabstract
\endtopmatter
\document
{\bf 1. Introduction}
\bigskip
The importance of antisymmetric tensor fields for string theory and for some supergravity models has been recognized for many years [1,2]. In
its original version, the corresponding configuration space of abelian gauge fields of rank $p$, denoted by $\Cal A^p$, consists of
differential $p$-forms on a manifold $M$ on which the abelian group $imd_{p-1}$ of exact $p$-forms acts by translation. To any gauge field
$A\in\Cal A^p$ a gauge invariant $(p+1)$-form field strength $F_A:=d_pA$ is associated which has vanishing magnetic flux by construction.
Examples are the neutral scalar field for $p=0$, the photon field for $p=1$, and in the low energy limit of type II string theory the
Kalb-Ramond field for $p=2$ as well as the Ramond-Ramond field, whose allowed rank $p$ depends on whether type IIA or type IIB string theory is
considered.\par

Recently, field configurations with non-trivial fluxes gained a prominent role in the dynamics of string theoretical models. From a
mathematical point of view neither the definition of the field strength $F_A$ as an exact differential form nor the interpretation of
$imd_{p-1}$ as the underlying symmetry group are appropriate concepts to describe such topologically non-trivial configurations.
\par
The so-called {\it generalized abelian gauge theories} provide a suitable mathematical framework for the description of such fields. In
principle, a generalized abelian gauge theory is a field theory whose gauge invariant field configurations belong to a (generalized)
differential cohomology group. In brief, (generalized) differential cohomology can be viewed as a specific combination of ordinary cohomology
or even generalized cohomology - meaning that the Eilenberg-Steenrod axioms except for the dimension axiom are satisfied - with the algebra of
closed differential forms on smooth manifolds. An exposition of the underlying concept and some applications in quantum field theory can be
found in [3]. The mathematical theory of this new topic has been introduced and elaborated in [4]. Differential cohomology groups with physical
relevance are in particular the group of Cheeger-Simons differential characters [5] and the (twisted) differential $K$-theory [3,4]. In this
respect we would like to mention the generalized Maxwell-theory, which has been analyzed in the Hamiltonian approach quite recently [6,7] and
the generalized Maxwell-Chern-Simons theory [8]. Furthermore the 2-form $B$-field in type II-superstring theory, the differences of the
so-called 3-form $C$-gauge fields of $M$-theory [9] and the gauge invariant classes of the Ramond-Ramond (RR) fields [10,11] can be naturally
interpreted in terms of differential cohomology.\par

The present paper is devoted to a study of covariant quantization of generalized abelian gauge theories on closed and compact manifolds of
dimension $n$. Our aims are to clarify the underlying geometrical structure, to construct an appropriate measure for the functional integral
and to determine the vacuum expectation values of physical observables. In our setting a generalized abelian gauge field is described by a
$p$-form gauge field $A\in\Cal A^p$ and a cohomology class $c\in H^{p+1}(M;\Bbb Z)$, which characterizes its topological type. Yet the
corresponding gauge group is the abelian group $\Omega _{\Bbb Z}^{p}(M;\Bbb R)$ of differential $p$-forms with integral periods, which acts
naturally on $\Cal A^p$. The space of inequivalent generalized gauge fields is identified with the abelian group of Cheeger-Simons differential
characters of rank $p$. Let us remind that a differential character $\hat u$ is a specific group homomorphism from singular $p$-cycles of $M$
to the 1-torus $\Bbb T^1$. A closed but non-exact differential $(p+1)$-form with integer periods, denoted by $\delta _1(\hat u)$, is assigned
to $\hat u$, which in physical terms is regarded as the field strength in the corresponding generalized abelian gauge theory.\par

We will prove explicitly that the space of gauge fields admits the structure of a non-trivial flat principal fiber bundle over the orbits which
are generated by the subgroup of smooth Cheeger-Simons differential characters. In physical terms this implies that it is impossible to obtain
a global smooth gauge fixing condition; the theory is said to suffer from gauge ambiguities. Topologically, the non-triviality of the bundle is
related to the free part of $H^p(M;\Bbb Z)$. Moreover, each orbit of the subgroup of smooth differential characters can be proven to be a
trivializable bundle over the torus $\Bbb T^{b_p}$, whose dimension is the $p$-th Betti number $b_p=dim H^p(M;\Bbb R)$ of $M$.\par

In order to elucidate the general concept in a concrete field theoretical model, we will consider the covariant quantization of the so-called
generalized $p$-form Maxwell theory. This model generalizes the (conventional) $p$-form Maxwell theory in so far as $F_A=d_pA$ is replaced by
the generalized field strength $\delta _1(\hat u)\in\Omega _{\Bbb Z}^{p+1}(M;\Bbb R)$ in the classical action $S_{inv}=\frac{1}{2}\int
_MF_A\wedge\star F_A$, where $\star$ denotes the Hodge star operator.\par

What could be a guiding principle for the construction of a functional integral measure for the generalized abelian gauge theory? Let us
briefly review the topologically trivial case: The partition function for the $p$-form Maxwell theory is defined as the functional integral

$$\Cal Z^{(p)}=\frac{1}{Vol(imd_{p-1})}\int _{\Cal A^{p}}vol_{\Cal A^{p}}\ e^{-S_{inv}},\tag1.1$$ over the field
space $\Cal A^p$. Here $vol_{\Cal A^{p}}$ is the formal volume form on $\Cal A^{p}$ and $Vol(imd_{p-1})$ denotes the infinite volume of the
gauge group $imd_{p-1}$. Due to the gauge invariance of the classical action $S_{inv}(A)$, the integrand in the numerator of (1.1) is constant
along the orbits of the gauge group, which have infinite measure. It is argued that by separating the divergent gauge dependent part from the
integrand and dividing by $Vol(imd_{p-1})$, which is infinite as well, the partition function (1.1) can be rendered finite. According to a
modified Faddeev-Popov approach [12-15], which takes the reducibility of the algebra of gauge transformations into account using the so called
"ghost-for-ghost" procedure, this separation can be provided by selecting gauge fixing conditions in all dimensions up to the rank of the gauge
fields. The resulting functional integral is evaluated over a global gauge fixing submanifold in $\Cal A^{p}$ with a weight factor given by the
Jacobians of the Faddeev-Popov operators associated with the given gauge fixing conditions. However, the Faddeev-Popov procedure fails if it is
impossible to fix the gauge globally.\par

An alternative way to quantize theories which are governed by degenerate action functionals has been introduced by Schwarz [16-18]. This method
of invariant integration relies on the reduction of the functional integral in (1.1) over $\Cal A^{p}$ to an integral over the corresponding
gauge orbit space $\Cal A^p/imd_{p-1}$ times the volume of the gauge group modified by a ghost-for-ghost contribution. In the infinite
dimensional case this extraction of the gauge group volume is ill-defined. However, Schwarz proposed to omit this infinite factor and take the
remaining functional integral over the gauge orbit space as the correct partition function of the theory.\par

One could raise the question if it is possible to include the gauge degrees of freedom in a reasonable way and to circumvent the gauge
ambiguities.\par

Instead of constructing a measure on the abelian group of Cheeger-Simons differential characters we search for a functional integral
formulation of generalized abelian gauge theory directly on $\Cal A^p$ including the different topological sectors in $H^{p+1}(M;\Bbb Z)$. At
first glance this approach seems to be of limited use due to the gauge ambiguities and the infinite dimensional gauge group. We propose a
functional integral measure that resolves these problems and provides a mathematically reasonable treatment of the gauge degrees of freedom.
For that we will apply a concept, which has been originally developed in the context of stochastic quantization of gauge theories [19,20]. In
principle, the construction of this functional integral measure relies on the following three steps: \roster\item A regularizing measure is
introduced for the gauge degrees of freedom yielding a finite volume of the gauge group. \item A prescription is given to select a unique
representative along each gauge orbit. The set of these representatives generates a gauge fixing submanifold. It is the correct region over
which the functional integral has to be taken. In topologically non-trivial situations, like we encounter in the generalized abelian theory,
the occurrence of gauge ambiguities prevents the existence of any smooth global gauge fixing. Hence the construction of gauge fixing
submanifolds can be done locally only.\item A family of measures is selected on $\Cal A^p$, whose domains of definition are determined by the
local gauge fixing submanifolds and which are integrable along the orbits of the gauge group. Finally these local measures are glued together
in such a way that the physical relevant objects become independent of the chosen gauge group regularization and of the particular way this
gluing was provided. Hence the problem of gauge ambiguities can be circumvented, guaranteeing the existence of an integrable partition function
on $\Cal A^p$.\endroster

This paper is structured as follows: In section 2 we will review briefly the concept of the Cheeger-Simons differential characters. The
geometrical structure of the configuration space of generalized gauge theories will be studied in section 3. Some of our results regarding the
geometrical structure of the inequivalent generalized gauge fields have been obtained in a different way in [6] and [21]. Section 4 is devoted
to the construction of the partition function and the vacuum expectation value (VEV) for generalized abelian gauge theories. We derive closed
expressions for the generalized $p$-form Maxwell theory. In the topologically trivial case, the partition function for the $p$-form Maxwell
theory can be recovered, yet the gauge group volume is sufficiently regularized. In section 5 the one-point and two-point functions are
explicitly computed showing non-trivial effects due to the topology of the gauge orbit space and the regularized gauge group volume. In section
6 the VEV is elaborated for circle-valued homomorphisms, which represent a natural class of gauge invariant observables. First, we study the
so-called smooth circle-valued homomorphisms, which can be characterized equivalently in terms of the Poincar\'e - Pontrjagin duality of
differential characters. Second, the Wilson operator, which is a multi-dimensional generalization of the Wilson operator for loops, is
considered for singular $p$-cycles of $M$. In both cases, we will find that the corresponding VEVs vanish unless specific topological
conditions are satisfied.
\bigskip\bigskip
{\bf 2. Setting the stage - Definition of the Cheeger-Simons differential characters}
\bigskip
In this section we want to recall the concept of the abelian group of differential characters which has been introduced by Cheeger and Simons
[5]. In the present paper, let $M$ be a $n$-dimensional closed, connected, oriented and compact Riemannian manifold and let us denote the
complex of smooth singular chains of $M$ with coefficients in $\Lambda =\Bbb Z,\Bbb R$ by $C_{\ast}(M;\Lambda)$ and its subcomplex of all
smooth singular cycles by $Z_{\ast}(M;\Lambda)$. Furthermore the boundary and coboundary operators will be denoted by $\partial$ and $\delta$,
respectively. There is a natural map between the complex of differential forms $\Omega ^{\ast}(M;\Bbb R)$ and $C^{\ast}(M;\Bbb R)$ given by
integration of differential forms over smooth singular chains. Let $q\colon\Bbb R\rightarrow\Bbb R/\Bbb Z$ be the reduction of $\Bbb R$ modulo
$\Bbb Z$, then there is an induced map

$$i\colon\Omega ^{\ast}(M;\Bbb R)\rightarrow C^{\ast}(M;\Bbb
R/\Bbb Z ),\quad i(A)(\sigma )=q(\int _{\sigma}A),\quad\sigma\in C_{\ast}(M;\Bbb Z).\tag2.1$$ We identify $\Bbb R/\Bbb Z$ with the 1-torus
$\Bbb T^1$ and take $q(.)=e^{2\pi\sqrt{-1}(.)}$. Let us introduce the abelian group
$$\Omega _{\Bbb Z}^p(M,\Bbb R)=\{A\in\Omega ^p(M;\Bbb R)
\vert\quad d\alpha =0,\quad \int _{\Sigma}\alpha \in\Bbb Z\quad \forall\Sigma\in Z_p(M;\Bbb Z)\}\tag2.2$$ of closed $\Bbb R$-valued
differential $p$-forms with integer periods.

\proclaim{Definition 2.1} The abelian group of Cheeger-Simons differential characters of degree $p$ is defined by
$$\hat{H}^p(M;\Bbb R/\Bbb Z)=\{ \hat{u}\in Hom(Z_p(M;\Bbb Z),\Bbb
R/\Bbb Z) |\quad \hat{u}\circ\partial \in i(\Omega _{\Bbb Z}^{p+1}(M,\Bbb R))\}.$$
\endproclaim
For any differential character $\hat u\in\hat{H}^p(M;\Bbb R/\Bbb Z)$ one can always find a cochain $u\in C^{p}(M;\Bbb R)$ whose restriction to
cycles is just $\hat u$. There exist a differential form with integer periods $F\in\Omega _{\Bbb Z}^{p+1}(M,\Bbb R)$ and a singular cochain
$c\in C^{p+1}(M;\Bbb Z)$, such that $\delta u=F-c$. Here $F$ satisfies the relation $(\hat{u})(\partial\sigma )=q(\int _{\sigma}F)$ for all
$\sigma\in C_{p+1}(M;\Bbb Z)$. Evidently, $c$ is a cocycle (i.e. $\delta c=0$), so that one can construct the following two homomorphisms on
$\hat H^p(M;\Bbb R/\Bbb Z)$, namely $\delta _1(\hat{u})=F$ and $\delta _2(\hat{u})=[c]$ .
\par
Set $R^{p+1}(M,\Bbb Z )=\{(F,c)\in\Omega _{\Bbb Z}^{p+1}(M,\Bbb R)\times H^{p+1}(M,\Bbb Z )|\quad r_{\ast}(c)=[F]\}$, where $r_{\ast}\colon
H^{p+1}(M,\Bbb Z )\rightarrow H^{p+1}(M,\Bbb R)$ is induced by the inclusion $\Bbb Z\hookrightarrow\Bbb R$ and $[F]$ is the cohomology class of
$F$ in $H^{p+1}(M;\Bbb R)$. The abelian group of Cheeger-Simons differential characters is characterized by the following result:

\proclaim{Theorem 2.2 [5]} The following sequences are exact
$$ \align & 0 @>>> H^p(M,\Bbb R/\Bbb Z) @> j_1 >> \hat{H}^p(M;\Bbb
R/\Bbb Z) @> \delta _1 >>\Omega _{\Bbb Z}^{p+1}(M,\Bbb R) @>>> 0 \\
& 0 @>>> \Omega ^p(M,\Bbb R)/\Omega _{\Bbb Z}^p(M,\Bbb R)
@> j_2 >> \hat{H}^p(M;\Bbb R/\Bbb Z )@> \delta _2 >> H^{p+1}(M,\Bbb Z ) @>>> 0 \\
& 0 @>>> H^p(M,\Bbb R )/r_{\ast}(H^p(M,\Bbb Z ))@> j_3 >> \hat{H}^p(M;\Bbb R/\Bbb Z )@> (\delta _1,\delta _2)
>>R^{p+1}(M,\Bbb Z ) @>>> 0, \endalign$$
where $j_1([v]):=v|_{Z_p(M;\Bbb Z)}$ is the restriction map and $j_2([A])(\Sigma ):=q(\int _{\Sigma}A)$ for $\Sigma\in Z_p(M;\Bbb Z)$. The
third sequence follows by combining the first two. \qed\endproclaim

Let us consider the long exact cohomology sequence
$$\cdots\rightarrow H^p(M;\Bbb R)@>q_{\ast}>>H^{p}(M;\Bbb R/\Bbb Z)@>\delta ^{\ast}>>H^{p+1}(M;\Bbb Z)@>r_{\ast}>>H^{p+1}(M;\Bbb R)@>>>\cdots\tag2.3$$
induced by the short exact sequence $0 @>>>\Bbb Z @>>>\Bbb R @>>>\Bbb T^1 @>>>0$, where $\delta ^{\ast}$ denotes the Bockstein operator, then
\roster\item $\delta _1|_{\Omega ^p(M,\Bbb R)/\Omega _{\Bbb Z}^p(M,\Bbb R)}:=\delta _1\circ j_2=d_p$ \item $\delta _2|_{H^p(M,\Bbb R/\Bbb
Z)}:=\delta _2\circ j_1=-\delta ^{\ast}$.\endroster

Additionally, the differential characters can be equipped with an associative, graded commutative ring structure
$$\ast\colon\hat{H}^{p_1}(M;\Bbb R/\Bbb
Z)\times\hat{H}^{p_2}(M;\Bbb R/\Bbb Z)\rightarrow\hat{H}^{p_1+p_1+1}(M;\Bbb R/\Bbb Z),\tag2.4$$ which according to [5] satisfies the following
relations (in "multiplicative" notation):\pagebreak\roster\item $(\hat u_1\ast\hat u_2)\ast\hat u_3=\hat u_1\ast(\hat u_2\ast\hat u_3)$\item
$\hat u_1\ast\hat u_2=(\hat u_2\ast\hat u_1)^{(-1)^{(p_1+1)(p_2+1)}}$, where $\hat u_1\in\hat H^{p_1}(M;\Bbb R/\Bbb Z)$ and $\hat u_2\in\hat
H^{p_2}(M;\Bbb R/\Bbb Z)$,\item $\delta _1(\hat u_1\ast\hat u_2)=\delta _1(\hat u_1)\wedge\delta _1(\hat u_2)$,\item $\delta _2(\hat
u_1\ast\hat u_2)=\delta _2(\hat u_1)\cup\delta _2(\hat u_2)$, where $\cup$ denotes the cup product in singular cohomology, \item $\hat u\ast
j_2([A])=(j_2([\delta _1(\hat u)\wedge A]))^{(-1)^{p_1+1}}$, with $\hat u\in\hat H^{p_1}(M;\Bbb R/\Bbb Z)$ and $[A]\in\Omega ^{p_2}(M;\Bbb
R)/\Omega _{\Bbb Z}^{p_2}(M;\Bbb R)$,\item $\hat u\ast j_1([v])=(j_1(\delta _2(\hat u)\cup [v]))^{(-1)^{p_1+1}}$, where $\hat u\in\hat
H^{p_1}(M;\Bbb R/\Bbb Z)$, $[v]\in H^{p_2}(M;\Bbb R/\Bbb Z)$.\endroster

Via theorem 2.2, the abelian group $\hat{H}^p(M;\Bbb R/\Bbb Z )$ carries a natural topology coming from the $C^{\infty}$-topology on
differential $(p+1)$-forms and the standard topology on $H^p(M,\Bbb R )/r_{\ast}(H^p(M,\Bbb Z ))$. With respect to this topology, $\ker{\delta
_2}$ is the connected component of the identity of this group and consists of the so called smooth differential characters, those which can be
represented by smooth differential $p$-forms.\par

Furthermore there exists a Poincar\'e-Pontrjagin duality of differential characters [22]: Let $[M]$ denote the fundamental cycle and let $<,>$
be the evaluation map. For each $p$ with $0\leq p<n$, the pairing

$$\hat H^{n-p-1}(M;\Bbb R/\Bbb Z)\times\hat H^{p}(M;\Bbb R/\Bbb Z)\rightarrow\Bbb T^1,\quad (\hat v,\hat u)\mapsto<\hat v\ast\hat u,[M]>\tag2.5$$
is non-degenerate. The induced injective map

$$\Cal D\colon\hat H^{n-p-1}(M;\Bbb R/\Bbb Z)\rightarrow\hat H^{p}(M;\Bbb R/\Bbb Z)^{\ast}:= Hom(\hat H^{p}(M;\Bbb R/\Bbb Z),\Bbb R/\Bbb Z),\quad
\Cal D(\hat v)(\hat u):=<\hat v\ast\hat u,[M]>\tag2.6$$ has a dense range in the group of continuous homomorphisms into the 1-torus $\Bbb T^1$.
This range consists exactly of those homomorphisms $\varrho\in\hat H^p(M;\Bbb R/\Bbb Z)^{\ast}$, so that there exists a $\tilde\varrho\in
\Omega _{\Bbb Z}^{n-p}(M;\Bbb R)$ with $\varrho (j_2[A])=q(\int _{M}A\wedge\tilde\varrho )$ for all $A\in\Omega ^p(M;\Bbb R)$. These
circle-valued homomorphisms are called smooth. For a detailed discussion on the various topological issues we refer to [22].
\bigskip\bigskip
{\bf 3. The geometrical structure of the configuration space of generalized abelian gauge fields}
\bigskip
Let $\Cal A^p:=\Omega ^p(M;\Bbb R)$ denote the configuration space of the $p$-form gauge fields on the manifold $M$. A generalized abelian
gauge field of rank $p$ is characterized by a $p$-form gauge field $A\in\Cal A^p$ and a cohomology class $c\in H^{p+1}(M;\Bbb Z)$. $\Cal A^p$
can be equipped with the flat Riemann structure

$$<A_1,A_2>=\int _M
\ A_1\wedge\star A_2,\tag3.1$$ where $\star$ is the Hodge star operator with respect to a fixed metric on $M$, satisfying $\star
^2=(-1)^{p(n-p)}$ on differential $p$-forms. The co-differential $d_p^{\ast}=(-1)^{n(p+1)+1}\star d_{n-p}\star\colon\Omega ^{p}(M;\Bbb
R)\rightarrow\Omega ^{p-1}(M;\Bbb R)$ gives rise to the Laplace operator $\Delta _p =d_{p+1}^{\ast}d_p+d_{p-1}d_{p}^{\ast}$. Let $Harm^p(M)$
denote the space of $\Bbb R$-valued harmonic $p$-forms and let $Harm^{p}(M)^{\bot}$ be its orthogonal complement. The Green\rq s operator [23]
is defined by

$$G_p\colon\Omega ^p(M;\Bbb R)\rightarrow Harm^{p}(M)^{\bot},\quad G_p=
(\Delta _p\vert _{Harm^{p}(M)^{\bot}})^{-1}\circ \Pi ^{Harm^{p}(M)^{\bot}},\tag3.2$$ where $\Pi ^{Harm^{p}(M)^{\bot}}$ is the projection of
$\Omega ^p(M;\Bbb R)$ onto $Harm^{p}(M)^{\bot}$. By construction $\Delta _p\circ G_p=G_p\circ\Delta _p =\Pi ^{Harm^{p}(M)^{\bot}}$.\par

The gauge group of the generalized abelian gauge theory is taken to be the abelian group $\Cal G^p:=\Omega ^{p-1}(M;\Bbb R)\times Harm_{\Bbb
Z}^p(M)$, where $Harm_{\Bbb Z}^p(M):=\Omega _{\Bbb Z}^{p}(M;\Bbb R)\cap Harm^p(M)$. There is a natural action of $\Cal G^p$ on $\Cal A^p$,
given by

$$(A,(\alpha _{p-1},\lambda _p ))\mapsto A\cdot (\alpha _{p-1},\lambda _p ):=A+d_{p-1}\alpha _{p-1} +\lambda _p ,
\qquad (\alpha _{p-1},\lambda _p )\in\Cal G^p.\tag3.3$$ Evidently this action is not free, possessing the abelian group $\ker{d_{p-1}}$ as its
isotropy group. Since there is only one orbit type, the abelian group of restricted gauge transformations $\Cal G_{\ast}^p:=\Cal
G^p/\ker{d_{p-1}}$ acts freely on $\Cal A^p$. Using the Hodge decomposition theorem it is easy to see that $\Cal G_{\ast}^p$ is isomorphic to
$imd_p^{\ast}\times Harm_{\Bbb Z}^p(M)$ provided by the map $[(\alpha _{p-1},\lambda _p )]\mapsto (d_p^{\ast}G_pd_{p-1}\alpha _{p-1},\lambda
_p)$.\par

Let $D_{n-p}\colon H^{n-p}(M;\Bbb Z)\rightarrow H_{p}(M;\Bbb Z)$, $D_{n-p}(\nu )=\nu\cap [M]$ be the Poincare duality isomorphism, where $\cap$
is the cap product [24]. Since the homology of $M$ is finitely generated with rank $b_p$ we shall choose a set of $p$-cycles $\gamma
_i^{(p)}\in Z_p(M,\Bbb Z)$, $i=1,\ldots ,b_p$, whose homology classes $[\gamma _i^{(p)}]$ provide a Betti basis, thus generating the free part
$H_p(M;\Bbb Z)/TorH_p(M;\Bbb Z)$ of $H_p(M;\Bbb Z)$, where $TorH_p(M;\Bbb Z)$ denotes the torsion part of the $p$-th homology group. According
to the following isomorphisms

$$H^{n-p}(M;\Bbb Z)/TorH^{n-p}(M;\Bbb Z)\cong H_{\Bbb Z}^{n-p}(M;\Bbb R) \cong Harm_{\Bbb
Z}^{n-p}(M),\tag3.4$$ a basis $(\rho _i^{(n-p)})_{i=1}^{b_{n-p}}\in Harm_{\Bbb Z}^{n-p}(M)$ can be selected from the basis
$(D_{n-p}^{-1}([\gamma _i^{(p)}]))_{i=1}^{b_{n-p}}$ for the free part of $H^{n-p}(M;\Bbb Z)$. Using the Poincare duality and the Universal
Coefficient Theorem it follows that the product

$$\split H^p(M;\Bbb Z)/TorH^p(M;\Bbb Z) &\times H^{n-p}(M;\Bbb Z)/TorH^{n-p}(M;\Bbb Z)
\rightarrow\Bbb Z\\  (\mu ,\nu )&\mapsto <\mu ,D_{n-p}(\nu )>=<\mu\cup\nu,[M]>,\endsplit\tag3.5$$ gives a perfect pairing [24]. Hence a basis
$(\rho _i^{(p)})_{i=1}^{b_p}\in Harm_{\Bbb Z}^p(M)$ can be adjusted so that

$$\int _{\gamma _j^{(p)}}\rho _i^{(p)} =\int _M\ \rho _i^{(p)}\wedge\rho _j^{(n-p)}=\delta
_{ij}\tag3.6$$ holds, implying $\int _{\gamma _j^{(p)}}\alpha =\int _M\alpha\wedge\rho _j^{(n-p)}$ for any $[\alpha ]\in H^p(M;\Bbb R)$.
Finally there exists an induced metric
$$h_{jk}^{(p)}=<\rho _{j}^{(p)},\rho _{k}^{(p)}>\tag3.7$$ on $Harm^p(M)$. For $p=0$, one has $\rho ^{(0)}=1$ and therefore $h^{(0)}=Vol (M)$.
There is an equivalent characterization of the restricted gauge group: \proclaim{Lemma 3.1} There exists an isomorphism $\Cal G_{\ast}^p\cong
\Omega_{\Bbb Z}^p(M;\Bbb R)$. \endproclaim \demo{Proof} Using the Hodge decomposition theorem one verifies easily that the map $\kappa
_p\colon\Omega_{\Bbb Z}^p(M;\Bbb R)\rightarrow\Cal G_{\ast}^p$, given by $\kappa _p(\beta _p):=(d_p^{\ast}G_p\beta _p,\sum _{j=1}^{b_p}(\int
_{\gamma _j^{(p)}}\beta _p)\rho _j^{(p)})$ is an isomorphism.\qed\enddemo

According to theorem 2.2, the gauge orbit space $\Cal M^p:=\Cal A^p/\Cal G_{\ast}^p$ can be identified with the abelian subgroup of smooth
differential characters. Let $\pi _{\Cal A^{p}}\colon\Cal A^{p}\rightarrow\Cal M^p$ with $\pi _{\Cal A^{p}}(A):=[A]$ be the natural projection.
Now we will state the two main results of this section:

\proclaim{Theorem 3.2} The abelian group $\Cal A^{p}$ admits the structure of a non-trivial flat principal $\Cal G_{\ast}^{p}$-bundle over
$\Cal M^p$ with projection $\pi _{\Cal A^{p}}$. This bundle is trivializable if $H_{\Bbb Z}^p(M;\Bbb R)=0$.\endproclaim

\demo{Proof} We are going to construct a bundle atlas explicitly. For this we have to define an open cover of the gauge orbit space and a family
of local sections. Let us consider the exact sequence of abelian groups

$$0\rightarrow\Bbb Z^{b_p}\rightarrow\Bbb R^{b_p}@>\exp{2\pi\sqrt{-1}(.)}>>\Bbb
T^{b_p}\rightarrow 1,\tag3.8$$ which geometrically describes the universal covering of the $b_p$-dimensional torus. An open cover $\Cal V^{(p)}$
of $\Bbb T^{b_p}$ is given by the following family of open sets

$$\Cal V^{(p)}=\{V_{a}^{(p)}\vert\quad
a:=(a_1,\ldots ,a_j,\ldots ,a_{b_p}),\quad a_{j}\in\Bbb Z_2=\{1,2\}\},\tag3.9$$ where each $V_a^{(p)}=V_{a_1}\times\cdots\times
V_{a_j}\times\cdots\times V_{a_{b_p}}$ is a open set in $\Bbb T^{b_p}$. Here $V_{1}= \Bbb T^1\backslash\{northern pole\}$ for $a_{j}=1$ and
$V_{2}= \Bbb T^1\backslash\{southern pole\}$ for $a_{j}=2$ provide an open cover for each 1-torus $\Bbb T^1$. Let us choose the following two
local sections of the universal covering $\Bbb R^1 \rightarrow \Bbb T^1$

$$\align & s_{a_{j}}(z)=\{ {\frac{1}{2\pi }\arccos |_{(0,\pi ]}\Re z\atop
\frac{1}{2\pi }\arccos |_{[\pi,2\pi )}\Re z }\qquad {\Im z\geq 0\atop \Im z<0},\quad a_{j}=1 \\
& s_{a_{j}}(z)=\{ {\frac{1}{2\pi }\arccos |_{(\pi ,2\pi ]}\Re z\atop \frac{1}{2\pi }\arccos |_{[2\pi,3\pi )}\Re z}\qquad {\Im z\leq 0\atop \Im
z>0}\quad a_{j}=2,\tag3.10\endalign$$ where $z=\Re z+\sqrt{-1}\Im z\in\Bbb T^1$. The locally constant transition functions
$g_{a_ja_j^{\prime}}^{\Bbb T^1}\colon V_{a_j}\cap V_{a_j^{\prime}}\rightarrow\Bbb Z$ are defined by

$$s_{a_j^{\prime}}(z_{j})=s_{a_{j }}(z_{j})+g_{a_{j}a_j^{\prime}}^{\Bbb T^{1}}(z_{j}).\tag3.11$$ Thus we can generate a family of $2^{b_p}$ local
sections $s_a\colon V_a\subset\Bbb T^{b_p}\rightarrow\Bbb R^{b_p}$, $s_a=(s_{a_1},\cdots ,s_{a_{b_p}})$ with transition functions
$g_{aa^{\prime}}^{\Bbb T^{b_p}}=(g_{a_1a_1^{\prime}}^{\Bbb T^1},\ldots ,g_{a_{b_p}a_{b_p}^{\prime}}^{\Bbb T^1})$. These local sections will be
used to construct a bundle atlas as follows: Let us define the following smooth surjective map $\pi _{\Cal M^{p}}\colon\Cal
M^{p}\rightarrow\Bbb T^{b_p}$ by

$$\pi _{\Cal M^{p}}([A])=(e^{2\pi\sqrt{-1}\int _M A\wedge\rho
_1^{(n-p)}},\ldots , e^{2\pi\sqrt{-1}\int _M A\wedge\rho _{b_p}^{(n-p)}}).\tag3.12$$ Then the family of open sets $U_a^{(p)}:=(\pi_{\Cal
M^{p}})^{-1}(V_a^{(p)})$ provides a finite open cover $\Cal U=\{U_a^{(p)}\}$ of the infinite dimensional manifold $\Cal M^{p}$. A bundle atlas
is given by $\varphi _{a}:U_{a}^{(p)}\times\Cal G_{\ast}^p\rightarrow(\pi _{\Cal A^{p}}^{-1}(U_a^{(p)})$, $\varphi _{a}([A],(\xi _{p-1}
,\lambda _p ))=A\cdot (\omega _{a}^{p}(A))^{-1}+d_{p-1}\xi _{p-1} +\lambda _p$. Its inverse reads $\varphi _{a}^{-1}(A)=([A],\omega
_{a}^{(p)}(A))$, where

$$\split &\omega _{a}^{(p)}\colon \pi _{\Cal A^{p}}^{-1}(U_{a}^{(p)})\rightarrow \Cal
G_{\ast}^{p},\quad \omega _{a}^{(p)}(A) = (d_{p}^{\ast}G_{p}A,\sum
_{j=1}^{b_p}\epsilon _{a_j}^{(p)}(A)\rho _j^{(p)}),\\
&\epsilon _{a_j}^{(p)}\colon \pi _{\Cal A^{p}}^{-1}(U_{a}^{(p)})\rightarrow\Bbb Z,\quad \epsilon _{a_{j}}^{(p)}(A) =\int _M (A\wedge\rho
_j^{(n-p)})- s_{a_{j}}(e^{2\pi\sqrt{-1}\int _M (A\wedge\rho _j^{(n-p)})}).\endsplit\tag3.13$$ The locally constant transitions functions $\hat
g_{aa^{\prime}}\colon U_a^{(p)}\cap U_{a^{\prime}}^{(p)}\rightarrow\Cal G_{\ast}^p$ are finally given by
$$\hat g_{aa^{\prime}}([A])=\left( 0,\sum _{j=1}^{b_p}g_{a_ja_j^{\prime}}^{\Bbb
T^{1}}(e^{2\pi\sqrt{-1}\int _M A\wedge\rho _j^{(n-p)}})\rho _j^{(p)}\right)\in \Cal G_{\ast}^p\tag3.14$$ showing explicitly that the
obstruction to trivialize the bundle belongs to $H_{\Bbb Z}^p(M;\Bbb R)$. With respect to the basis $(\rho _{j}^{(p)})_{j=1}^{b_p}$ the
orthogonal projector onto $Harm^p(M)$ becomes

$$\Pi ^{Harm^p(M)}(A)=\sum _{j,k=1}^{b_p}(h_{jk}^{(p)})^{-1}<A,
\rho _{j}^{(p)}>\rho _{k}^{(p)},\ \forall A\in\Cal A^p.\tag3.15$$ From $\epsilon _{a_j}^{(p)}(A\cdot (\xi _{p-1},\lambda _p ))=\epsilon
_{a_j}^{(p)}(A)+\int _{\gamma _j^{(p)}}\lambda _p$ and $\Pi ^{Harm^p(M)}(\lambda _p )=\sum _{j=1}^{b_p}(\int _{\gamma _j^{(p)}}\lambda _p)\rho
_j^{(p)}$ one finally gets $\omega _a^{(p)}(A\cdot (\xi _{p-1},\lambda _p ))=\omega _a^{(p)}(A)\cdot (\xi _{p-1},\lambda _p )$. \qed\enddemo

Thus we have shown that it is impossible to choose a global gauge fixing condition, if the free part of $H^{p}(M;\Bbb Z)$ is non-vanishing.
Furthermore the subgroup of smooth differential characters has the following bundle structure:

\proclaim{Theorem 3.3} The manifold $\Cal M^p$ admits the structure of a trivializable vector bundle over $\Bbb T^{b_p}$ with typical fiber $im
d_{p+1}^{\ast}$ and projection $\pi _{\Cal M^p}$.
\endproclaim

\demo{Proof} A bundle atlas is provided by the diffeomorphisms $\chi _a \colon V_a^{(p)}\times im d_{p+1}^{\ast}\rightarrow \Cal M^p$

$$\split & \chi _{a}(z_1,\ldots ,z_{b_p},\tau _p)=[\sum _{j=1}^{b_p} s_{a_j}(z_j)\rho _j^{(p)}+\tau _p ]\\ &\chi _a^{-1}([A])
=([A], d_{p+1}^{\ast}G_{p+1}d_pA)\endsplit\tag3.16$$ On each fiber $(\pi _{\Cal M^p})^{-1}(z_1,\ldots ,z_{b_p})$ there is a unique structure of
a real vector space induced by the bundle chart $\chi _{a}$, giving rise to a natural vector bundle structure on $\Cal M^p$. \qed\enddemo

The abelian group $\Cal M^p$ acts freely on $\hat{H}^p(M;\Bbb R/\Bbb Z)$ by $\hat v\mapsto\hat v\cdot j_2([A])$ for $[A]\in\Cal M^p$. Let us
denote by $\Cal M_{\hat u}^p$ the corresponding orbit (i.e. the $\Cal M^p$-torsor) through $\hat u\in\hat{H}^p(M;\Bbb R/\Bbb Z)$. Then the
homomorphism $\delta _2$ is constant on $\Cal M_{\hat u}^p$. We define the projections $\pi _{\Cal A^{p}}^{\hat u}\colon\Cal A^p\rightarrow
\Cal M_{\hat u}^p$, by $\pi _{\Cal A^p}^{\hat u}(A)=\hat u\cdot j_2([A])$ and $\pi _{\Cal M_{\hat u}^p}\colon\Cal M_{\hat u}^p\rightarrow\Bbb
T^{b_p}$ by $\pi _{\Cal M_{\hat u}^p}(\hat{v})=\pi _{\Cal M^{p}}([A])$, where $\hat v=\hat u\cdot j_2([A)]$.

\proclaim{Corollary 3.4} For any fixed $\hat{u}\in\hat{H}^p(M;\Bbb R/\Bbb Z)$, the following holds:\roster\item The abelian group $\Cal A^{p}$
admits the structure of a non-trivial flat principal $\Cal G_{\ast}^{p}$-bundle over $\Cal M_{\hat u}^p$ with projection $\pi _{\Cal
A^{p}}^{\hat u}$. \item The manifold $\Cal M_{\hat u}^p$ admits the structure of a trivializable vector bundle over $\Bbb T^{b_p}$ with typical
fiber $im d_{p+1}^{\ast}$ and projection $\pi _{\Cal M_{\hat u}^p}$.\endroster
\endproclaim

\demo{Proof} Let $\Cal U(\hat{u})=\{U_a^{\hat{u}}\}$ be the induced open cover of $\Cal M_{\hat u}^p$, where $U_a^{\hat{u}}=\{\hat{u}\cdot
j_2([A])\in\hat{H}^p(M;\Bbb R/\Bbb Z)|[A]\in U_a^{(p)}\}$. A bundle atlas for $\Cal A^{p}$ is given by $\varphi
_{a}^{\hat{u}}:U_{a}^{\hat{u}}\times\Cal G_{\ast}^p\rightarrow(\pi _{\Cal A^{p}}^{\hat u})^{-1}(U_a^{\hat{u}})$,

$$\varphi _{a}^{\hat{u}}(\hat v,(\xi _{p-1} ,\lambda _p ))=A\cdot (\omega _{a}^{p}(A))^{-1}+d_{p-1}\xi _{p-1} +\lambda _p,\quad
(\varphi _{a}^{\hat{u}})^{-1}(A)=(\hat u\cdot j_2([A]),\omega _{a}^{(p)}(A)),\tag3.17$$ for $\hat v=\hat u\cdot j_2([A])$. A bundle atlas for
$\Cal M_{\hat u}^p$ is provided by the diffeomorphisms $\chi _a^{\hat{u}} \colon V_a^{(p)}\times im d_{p+1}^{\ast}\rightarrow \Cal M_{\hat
u}^p$ namely

$$\chi _{a}^{\hat{u}}(z_1,\ldots ,z_{b_p},\tau _p)=\hat{u}\cdot j_2(\chi _a(z_1,\ldots ,z_{b_p},\tau _p)),\quad (\chi
_a^{\hat{u}})^{-1}(\hat v) =\left( \pi _{\Cal M^p}^{\hat u} (\hat{v}), d_{p+1}^{\ast}G_{p+1}(\delta _1(\hat{v})-\delta
_1(\hat{u}))\right).\tag3.18$$

On each fiber $(\pi _{\Cal M_{\hat u}^p})^{-1}(z_1,\ldots ,z_{b_p})$ there is a unique structure of a real vector space induced by the bundle
chart $\chi _{a}^{\hat{u}}$, giving rise to a natural vector bundle structure on $\Cal M_{\hat u}^p$. \qed\enddemo

In fact, for any fixed $\hat u$, the bundle $\Cal A^p\rightarrow\Cal M_{\hat u}^p$ can be regarded equivalently as pull-back of $\Cal
A^p\rightarrow\Cal M^p$ via the diffeomorphism $\Upsilon _{\hat u}\colon\Cal M_{\hat u}^p\rightarrow \Cal M^p$ defined by $\Upsilon _{\hat
u}(\hat v):=[A]$, where $\hat v=\hat u\cdot j_2([A])$. Fields belonging to $\Cal M_{\hat u}^p$ may be called inequivalent generalized gauge
fields with topological type $\delta _2(\hat u)\in H^{p+1}(M;\Bbb Z)$.

\proclaim{Corollary 3.5} For any $\hat u$, the manifold $\Cal M_{\hat u}^p$ possesses the following topological information:
$$\split & H^k(\Cal M_{\hat u}^p,\Bbb Z)=H^k(\Bbb T^{b_p},\Bbb
Z)=\Bbb Z^{\binom {b_p}k}\\ & \pi _1(\Cal M_{\hat u}^p)=\pi _0(\Cal G_{\ast}^p)=\Bbb Z^{b_p}\\ & \pi_k(\Cal M_{\hat u}^p)=\pi _{k-1}(\Cal
G_{\ast}^p)=0\quad k\geqq 2\endsplit\tag3.19$$\qed\endproclaim

There is one remaining question: How does the choice of the fixed differential character $\hat u$ affect the vector bundle structure of $\Cal
M_{\hat u}^p$?

\proclaim{Proposition 3.6} For any two fixed differential characters $\hat u_1, \hat u_2\in\hat{H}^p(M;\Bbb R/\Bbb Z)$ one gets:

\roster\item $\Cal A^p@>\pi _{\Cal A^p}^{\hat u_1} >>\Cal M_{\hat u_1}^p$ and $\Cal A^p@>\pi _{\Cal A^p}^{\hat u_2} >>\Cal M_{\hat u_2}^p$ are
isomorphic as principal $\Cal G_{\ast}^p$-bundles,\item $\Cal M_{\hat u_1}^p@>\pi _{\Cal M_{\hat u_1}^p} >>\Bbb T^{b_p}$ and $\Cal M_{\hat
u_2}^p@>\pi _{\Cal M_{\hat u_2}^p}>>\Bbb T^{b_p}$ are isomorphic with respect to their vector bundle structures.\endroster\endproclaim

\demo{Proof} Let us consider the invertible map $\Upsilon ^{\hat u_1,\hat u_2}\colon\Cal M_{\hat u_1}^p\rightarrow\Cal M_{\hat u_2}^p$, with
$\Upsilon ^{\hat u_1,\hat u_2}(\hat v):=\hat u_2\cdot j_2([A])$, for $\hat v=\hat u_1\cdot j_2([A])$. Then $\Upsilon ^{\hat u_1,\hat u_2}$ fits
into the following commutative diagram of bundles

$$\CD \Cal A^p @= \Cal A^p \\ @V\pi _{\Cal A^p}^{\hat u_1}VV @VV\pi _{\Cal A^p}^{\hat u_2}V \\
\Cal M_{\hat u_1}^p @>\Upsilon ^{\hat u_1,\hat u_2}>> \Cal M_{\hat u_2}^p \\
@V\pi _{\Cal M_{\hat u_1}^p}VV @VV\pi _{\Cal M_{\hat u_2}^p}V \\ \Bbb T^{b_p} @= \Bbb T^{b_p},\endCD\tag3.20$$ proving that both the principal
$\Cal G_{\ast}^p$-bundle structure as well as the vector bundle structure are compatible.\qed\enddemo
\bigskip\bigskip
{\bf 4. The partition function for the generalized abelian gauge theory}
\bigskip
In this section we want to introduce a well-defined partition function for generalized abelian gauge theories. Furthermore we will study the
vacuum expectation value (VEV) of gauge invariant quantities.\par

A {\it gauge invariant observable} is any complex-valued (continuous) function on $\hat H^p(M;\Bbb R/\Bbb Z)$. For every arbitrary but fixed
differential character $\hat{u}\in\hat{H}^p(M;\Bbb R/\Bbb Z)$ and any gauge invariant observable $f$, there is an induced map $f^{\hat
u}\colon\Cal M^p\rightarrow\Bbb C$ defined by $f^{\hat u}([A]):=f(\hat u\cdot j_2([A])$. \par

Our proposal for constructing an appropriate functional integral takes up a method which has been applied successfully to the stochastic
quantization of Yang-Mills theory [19,20]. The main idea is to define an integrable partition function on the original field space $\Cal A^p$
by implementing a regularization of the volume of the gauge group, however, without affecting the VEV of gauge invariant observables. It will
be shown that this requirement is related to the problem of gauge fixing.\par

Now we follow the three steps, which have been indicated in section 1. Let $S_{reg}^{(p)}$ denote a real-valued function on $\Cal G_{\ast}^{p}$
such that the volume of the restricted gauge group

$$Vol(\Cal G_{\ast}^{p};e^{-S_{reg}^{(p)}}):=\int _{\Cal G_{\ast}^{p}}vol_{\Cal G_{\ast}^{p}}\ e^{-S_{reg}^{(p)}}=
\sum\limits _{\lambda _p\in Harm_{\Bbb Z}^p(M)}\int _{imd_p^{\ast}}\Cal D\tau _{p-1}\ e^{-S_{reg}^{(p)}(\tau _{p-1},\lambda _p)}\tag4.1$$
becomes finite. Analogously we assume the existence of a functional $Q^{(p)}$, which renders the volume

$$Vol(\Cal G^p;Q^{(p)})=\int _{\Cal G^p}vol_{\Cal G^p}\ Q^{(p)}=\sum\limits _{\lambda _p\in Harm_{\Bbb Z}^p(M)}\int
_{\Omega ^{p-1}(M)}\Cal D\alpha _{p-1}\ Q^{(p)}(\alpha _{p-1},\lambda _p)\tag4.2$$ of the total gauge group $\Cal G^{p}$ finite. Below we will
give explicit expressions for $S_{reg}^{(p)}$ and $Q^{(p)}$. For $p=0$, $\Cal G^p=\Cal G_{\ast}^p=\Bbb Z$ and thus one can set
$Q^{(0)}=S_{reg}^{(0)}$.\par

Let $S_{inv}(\hat u)$ denote the gauge invariant classical action of the generalized abelian gauge theory. We introduce the following family of
locally defined volume forms on the open sets $\pi _{\Cal A^p}^{-1}(U_a^{(p)})$

$$\Xi _a(\hat u_0^c):=vol_{\Cal A^p}|_{\pi _{\Cal A^p}^{-1}(U_a^{(p)})}e^{-S_{inv}^{\hat u_0^c}-
(\omega _a^{(p)})^{\ast}S_{reg}^{(p)}},\tag4.3$$ where $vol_{\Cal A^p}$ denotes the (formal) volume form on $\Cal A^p$ induced by (3.1). Each
of these local volume forms is damped along the gauge orbits and yields a VEV of gauge invariant observables which is independent of the
explicit form for $S_{reg}^{(p)}$. Due to the gauge ambiguities, $\Xi _a$ cannot be extended to a global differential form on the whole
configuration space $\Cal A^p$ in a natural way. However, in order to obtain a global integrand for the functional integral, we will glue these
local forms together using a partition of unity $\{ \hat g_a^{(p)} \}$ on $\Cal M^p$ subordinate to the open cover $\Cal U$. In fact, let
$\{\tilde{g}_{a_{j}}|a_{j}\in\Bbb Z_2,\}$ be a partition of unity on the $j$-th 1-torus $\Bbb T^1$ within $\Bbb T^{b_p}$ subordinate to the
open cover $\{V_{a_{j}}\}$. By means of the projection $q_{j}\colon\Bbb T^{b_p}@>>>\Bbb T^1$, $q_{j}(z_{1},\ldots ,z_{j},\ldots
,z_{b_p})=z_{j}$ onto the $j$-th one-torus $\Bbb T^1$, the functions $g_{a}^{(p)}:=\prod _{j=1}^{b_p}q_{j}^{\ast}\tilde g_{a_{j}}$ induce a
partition of unity of $\Bbb T^{b_p}$ subordinate to $\{V_{a}^{(p)}\}$. Finally $\hat g_a^{(p)}:=(\pi _{\Cal M^p})^{\ast}g_a^{(p)}$ is the
sought-after partition of unity subordinate to the open cover $\Cal U$ of the gauge orbit space $\Cal M^p$.\par

\proclaim{Lemma 4.1} To every $c\in H^{p+1}(M;\Bbb Z)$, one can assign a differential character $\hat{u}_0^c\in\hat{H}^p(M;\Bbb R/\Bbb Z)$ -
hereafter called background differential character - satisfying $\delta _2(\hat u_0^c)=c$ in such a way that $d_{p+1}^{\ast}\delta
_1(\hat{u}_0^c)=0$. A background differential character is unique up to the those smooth differential characters which are induced by closed
differential $p$-forms.\endproclaim \demo{Proof} Let $\hat{u}_0^{\prime}\in\hat{H}^p(M;\Bbb R/\Bbb Z)$. Then the modified differential
character $\hat{u}_0=\hat{u}_0^{\prime}\cdot j_2([-G_pd_{p+1}^{\ast}\delta _1(\hat{u}_0^{\prime})])$ satisfies the requested equation.
Moreover, $\delta _2(\hat u_0^c)=\delta _2(\hat u_0^{\prime})=c$. A direct calculation shows that any other background differential character
relates to $\hat u_c^0$ by $\hat u_c^0\cdot j_2([B])$, where $d_pB=0$. \qed\enddemo

Let $\{\hat u_0^c|c\in H^{p+1}(M;\Bbb Z)\}$ be a family of background differential characters. We introduce the following functional on gauge
invariant observables $f$, by

$$\Cal I^{(p)}(f):=\sum _{c\in H^{p+1}(M,\Bbb Z)}\frac{1}{Vol(\Cal G^p;Q^{(p)})}
\int _{\Cal A^p}\sum _{a\in\Bbb Z_2^{b_p}}\ (\pi _{\Cal A^p}^{\ast}\hat g_a^{(p)})\ \Xi _a(\hat u_0^c) (\pi _{\Cal A^p}^{\ast}f^{\hat
u_0^c}),\tag4.4$$ where $a=(a_{1},\ldots ,a_{b_p})\in\Bbb Z_2^{b_p}$ is a multi-index. Here the finite volume of the gauge group $\Cal G^p$ is
factored out in order to eliminate all unphysical degrees of freedom. Now we can state the main definition of this paper:

\proclaim{Definition 4.2} The partition function of the generalized abelian gauge theory is defined by

$$\Cal Z^{(p)}:=\Cal I^{(p)}(1).\tag4.5$$ The VEV of a gauge invariant observable $f$ is defined by

$$\Cal E^{(p)}(f):=\frac{\Cal I^{(p)}(f)}{\Cal Z^{(p)}}.\tag4.6$$
\endproclaim

In order to make our concept explicit, we will consider the {\it generalized $p$-form Maxwell theory}. This field theory is governed by the
classical action

$$S_{inv}(\hat u)=\frac{1}{2}\|\delta _1(\hat u)\|^2=\frac{1}{2}\int _M\delta _1(\hat u)\wedge\star\delta _1(\hat u).\tag4.7$$
\subheading{Examples}\roster\item If $H^{p+1}(M;\Bbb Z)=0$, every differential character $\hat u$ is smooth; i.e. $\hat{u}=j_2([A])$ for some
$A\in\Cal A^p$. Since $\delta _1(\hat u)=d_pA=F_A$, (4.7) reduces to the classical action for the $p$-form Maxwell theory. \item In the case
$p=0$, one has $\hat H^0(M;\Bbb R/\Bbb Z)\cong C^{\infty}(M;\Bbb T^1)$. It follows from theorem 2.2 that $\delta _1(\hat u)=\hat
u^{\ast}\vartheta$, where $\vartheta $ is the Maurer Cartan form on $\Bbb T^1$. So we recover the action $S_{inv}(\hat u)=\frac{1}{2}\|\hat
u^{\ast}\vartheta\|^2$ for circle-valued scalar fields on $M$.\endroster

The explicit calculation of (4.4) will be done in five steps: The integration over $\Cal A^{p}$ is separated into an integration over the base
manifold $\Bbb T^{b_p}$ and an integration over the fiber $\Cal G_{\ast}^{p}\times imd_{p+1}^{\ast}$. Second, the division by the gauge group
volume is carried out. In the third step it is shown that (4.4) does not depend on the concrete choice for the background connections. The
summation over the different topological sectors is performed in step four. Finally we construct realizations for $S_{reg}^{(p)}$ and $Q^{(p)}$
in order to obtain an explicit expression for the partition function in the original configuration space $\Cal A^p$.

\subheading {Split of the integral over $\Cal A^p$} Let us introduce the following family of local diffeomorphisms, $\psi _a=\varphi _a\circ
(\chi _a\times\Bbb I)\colon V_a^{(p)}\times imd_{p+1}^{\ast}\times\Cal G_{\ast}^{p}\rightarrow (\pi _{\Cal M^{p}}\circ\pi_{\Cal
A^{p}})^{-1}(V_a^{(p)})$, using the results of theorems 3.2 and 3.3. The induced local metrics are

$$\split &((\psi _a)^{\ast}<,>)_{(z_1,\ldots ,z_{b_p},\tau _{p},(\tau _{p-1},\lambda _p))}
\left( (w_1^{1},\ldots ,w_{b_p}^{1},u_{p}^{1},(\upsilon _{p-1}^{1},0)),(w_1^{2},\ldots , w_{b_p}^{2},u_{p}^{2},
(\upsilon _{p-1}^{2},0))\right)=\\
& =\frac{1}{(2\pi )^2}\sum \Sb j=1\\ k=1\endSb ^{b_p}\overline{\vartheta _{z_{j}}(w_{j}^{1})}\vartheta
_{z_{k}}(w_{k}^{2})h_{jk}^{(p)}+<u_{p}^{1},u_{p}^{2}>+<\upsilon _{p-1}^{1},\Delta _{p-1}|_{imd_{p}^{\ast}}\upsilon
_{p-1}^{2}>,\endsplit\tag4.8$$ where $z_j\in\Bbb T^1$, $w_j\in T_{z_j}\Bbb T^{1}$ for $j=1,\ldots b_p$, $u_{p}\in T_{\tau
_{p}}imd_{p+1}^{\ast}$, $(\upsilon _{p-1},0))\in T_{(\tau _{p-1},\lambda _p)}\Cal G_{\ast}^{p}$ and $\bar\vartheta$ denotes the complex
conjugate of the Maurer Cartan form on $\Bbb T^1$. The corresponding volume form is given by

$$\psi _a^{\ast}vol_{\Cal A^{p}}=\cases \frac{1}{2\pi}(Vol(M))^{1/2}\ vol_{\Bbb T^{1}}\vert _{V_a^{(0)}}\wedge
vol_{imd_{1}^{\ast}} &\text{if $p=0$} \\ \frac{1}{(2\pi )^{b_p}}(\det{h^{(p)}})^{1/2}\det{(\Delta _{p-1}|_{imd_{p}^{\ast}})}^{1/2} vol_{\Bbb
T^{b_p}}\vert _{V_a^{(p)}}\wedge vol_{imd_{p+1}^{\ast}}\wedge vol_{\Cal G_{\ast}^p} &\text{if $p\neq 0$,}\endcases\tag4.9$$ where $vol_{\Bbb
T^{b_p}}=(\sqrt{-1})^{-b_p}q_{1}^{\ast}\vartheta\wedge\ldots \wedge q_{b_p}^{\ast}\vartheta $ is the induced volume form on $\Bbb T^{b_p}$ and
$vol_{imd_{p+1}^{\ast}}$ denotes the flat metric on $imd_{p+1}^{\ast}$, which is induced by (3.1). Since the bundle $\Cal A^p@>>>\Cal M^p$ is
flat, the local volume forms in (4.3) can be glued together to yield a global volume form on the product space $\Bbb T^{b_p}\times
imd_{p+1}^{\ast}\times\Cal G_{\ast}^{p}$. Let us remark that the determinants arising in (4.9) are understood in terms of zeta-regularization
[16]: For any non-negative self-adjoint elliptic operator $\Cal B$ its regularized determinant can be defined by

$$\det\Cal B =\exp{\left( -\frac{d}{ds}|_{s=0}\zeta (s|\Cal B)\right)},\tag4.10$$ where $\zeta (s|\Cal B)$ is the zeta-function of
the operator $\Cal B$, namely

$$\zeta (s|\Cal B)=\sum _{\nu _j\neq 0}\nu _j ^{-s}=\frac{1}{\Gamma (s)}\int _0^{\infty}t^{s-1}
Tr(e^{-t\Cal B}-\Pi ^{\Cal B })dt.\tag4.11$$ Here $\nu _j$ are the non-vanishing eigenvalues of $\Cal B$ and $\Pi ^{\Cal B}$ is the othogonal
projector onto the kernel of $\Cal B$. The $\zeta$-function is analytic at the origin and possesses a meromorphic extension over $\Bbb C$.\par

Let $f$ be an arbitrary gauge invariant observable. Due to its gauge invariance, $\tilde {f}_a^{\hat u_0^c}:=(\chi _{a})^{\ast}f^{\hat{u}_0^c}$
can be extended to a globally defined function $\tilde {f}^{\hat u_0^c}$ on $\Bbb T^{b_p}\times imd_{p+1}^{\ast}$.\par

Let $e_{(m_{1},\ldots ,m_{b_p})}(z_{1},\ldots ,z_{b_p}):=z_{1}^{m_{1}}\cdots z_{b_p}^{m_{b_p}}$ be an orthonormal basis of $L^2(\Bbb
T^{b_p};\Bbb C)$ with respect to the inner product $\ll f_1,f_2 \gg:=\frac{1}{(2\pi )^{b_p}}\int _{\Bbb T^{b_p}}vol_{\Bbb
T^{b_p}}\bar{f}_1f_2$, where $z_j\in\Bbb T^1$ and $m_{j}\in\Bbb Z$. So $\tilde {f}^{\hat u_0^c}(.,\tau _p)$ can be rewritten in terms of a
Fourier series expansion as

$$\tilde{f}^{\hat u_0^c}(z_{1},\ldots ,z_{b_p},\tau _p )=\sum _{m_{1}\in\Bbb Z}\cdots\sum _{m_{b_p}\in\Bbb Z}\tilde{f}^{\hat u_0^c}_{(m_{1},\ldots
,m_{b_p})}(\tau _p)\ z_{1}^{m_{1}}\cdots z_{b_p}^{m_{b_p}}, \tag4.12$$ where $z_j=e^{2\pi\sqrt{-1}w_{j}}$ and the Fourier coefficients read

$$\tilde{f}^{\hat u_0^c}_{(m_{1},\ldots ,m_{b_p})}(\tau _p)=\int\limits _{0}^{1}\cdots
\int\limits _{0}^{1}dw_{1}\ldots dw_{b_p} \tilde{f}^{\hat u_0^c}(e^{2\pi\sqrt{-1}w_{1}},\ldots ,e^{2\pi\sqrt{-1}w_{b_p}},\tau
_p)e^{-2\pi\sqrt{-1}\sum\limits _{j=1}^{b_p}m_{j}w_{j}}.\tag4.13$$ Using (4.9) and (4.13) the non-normalized VEV of gauge invariant observables
(4.4) can be rewritten in the form

$$\Cal I^{(0)}(f)=Vol(M)^{1/2}\sum _{c\in H^{1}(M;\Bbb Z)}e^{-\frac{1}{2}||(\hat u_0^c)^{\ast}\vartheta||^2}
\int _{imd_{1}^{\ast}}\Cal D\tau _0\ \tilde{f}^{\hat u_0^c}_{(0)}(\tau _0)\ e^{-\frac{1}{2}<\tau _0,\Delta _0|_{imd_{1}^{\ast}}\tau
_0>},\tag4.14$$ for $p=0$ and

$$\split\Cal I^{(p)}(f) &=(\det{h^{(p)}})^{1/2}(\det{\Delta _{p-1}|_{imd_p^{\ast}}})^{1/2} \frac{Vol(\Cal G_{\ast}^{p};e^{-S_{reg}^{(p)}})}{Vol(\Cal
G^p;Q^{(p)})} \sum _{c\in H^{p+1}(M;\Bbb Z)}e^{-\frac{1}{2}||\delta _1(\hat u_0^c)||^2} \\ &\times\int _{imd_{p+1}^{\ast}}\Cal D\tau _p\
\tilde{f}^{\hat u_0^c}_{(0,\ldots ,0)}(\tau _p)\ e^{-\frac{1}{2}<\tau _p,\Delta _p|_{imd_{p+1}^{\ast}}\tau _p>},\endsplit\tag4.15$$ for $p\neq
0$.

\subheading {Geometry and regularization of the gauge group} The next task is to calculate the factor $\frac{Vol(\Cal
G_{\ast}^{p};e^{-S_{reg}^{(p)}})}{Vol(\Cal G^p;Q^{(p)})}$ in (4.15). We will prove that this fraction equals the inverse of the regularized
volume of the isotropy group of the action of $\Cal G^p$ on $\Cal A^p$. Let us consider the following two families of abelian groups $\Cal
G^k:=\Omega ^{k-1}(M)\times Harm_{\Bbb Z}^k(M)$ and $\Cal G_{\ast}^k:=imd_{k}^{\ast}\times Harm_{\Bbb Z}^k(M)$ with $k=1,\ldots ,p-1$.

\proclaim{Proposition 4.3} For any $k$, $1\leq k\leq p$, $\ker{d_{k-1}}$ admits the structure of a non-trivial flat principal $\Cal
G_{\ast}^{k-1}$-bundle over $\Bbb T^{b_{k-1}}$ with projection $\tilde\pi _{k-1}(\beta _{k-1})=(e^{2\pi\sqrt{-1}\int _{\gamma _1^{(k-1)}}\beta
_{k-1}},\ldots ,e^{2\pi\sqrt{-1}\int _{\gamma _{b_{k-1}}^{(k-1)}}\beta _{k-1}})$.\endproclaim

\demo{Proof} This results follows directly from theorems 3.2 and 3.3. Like in (3.9) we choose the open cover $\Cal V^{(k-1)}$ of $\Bbb
T^{b_{k-1}}$ and the following bundle atlas

$$\split & \tilde\psi _a^{(k-1)} \colon V_a^{(k-1)}\times\Cal G_{\ast}^{k-1}\rightarrow \tilde\pi _k^{-1}(V_a^{(k-1)})\\ &
\tilde\psi _a^{(k-1)}(z_1,\ldots ,z_{b_{k-1}},\xi _{k-2},\lambda _{k-1})=(\sum _{j=1}^{b_{k-1}}s_{a_j}(z_j)\rho _j^{(k-1)}+d_{k-2}\xi
_{k-2},\lambda _{k-1})
\\ &(\tilde\psi _a^{(k-1)})^{-1}(\beta _{k-1})=(\tilde\pi _{k-1}(\beta _{k-1}),\omega _a^{(k-1)}(\beta _{k-1})),\endsplit\tag4.16$$ where
$\omega _a^{(k-1)}$ is taken as in (3.13) yet $p=k-1$. \qed\enddemo

For each $k=1,\ldots ,p$, let us take a regularizing function $S_{reg}^{(k)}$ for the abelian group $\Cal G_{\ast}^{k}$, such that $Vol(\Cal
G_{\ast}^{k};e^{-S_{reg}^{(k)}})$ becomes finite. Then $\tilde Q^{(k-1)}:=\sum\limits _{a\in\Bbb Z_2^{b_{k-1}}}(\tilde\pi
_{k-1}^{\ast}g_a^{(k-1)})e^{-(\omega _a^{(k-1)})^{\ast}S_{reg}^{(k-1)}}$ provides a regularization for the isotropy group $\ker{d_{k-1}}$, since

$$\split Vol(\ker{d_{k-1}};\tilde Q^{(k-1)})&=\int _{\ker{d_{k-1}}}vol_{\ker{d_{k-1}}}\tilde Q^{(k-1)}\\
&=(\det{h^{(k-1)}})^{1/2}(\det{\Delta _{k-2}|_{imd_{k-1}^{\ast}}})^{1/2}Vol(\Cal G_{\ast}^{k-1};e^{S_{reg}^{(k-1)}}),\endsplit\tag4.17$$ is
finite. Here $vol_{\ker{d_{k-1}}}$ is the volume form on $\ker{d_{k-1}}$ induced by the flat metric (3.1).

\proclaim{Proposition 4.4} For any fixed $k$, $1\leq k\leq p$, the abelian group $\Cal G^k$ admits the structure of a non-trivial flat
principal $\Cal G_{\ast}^{k-1}$-bundle over the base manifold $\Bbb T^{b_{k-1}}\times\Cal G_{\ast}^k$ with projection $\pi _{k-1}(\alpha
_{k-1},\lambda _k):=(\pi _{\Cal M^{k}}\circ\pi _{\Cal A^{k}}(\alpha _{k-1}),d_k^{\ast}G_pd_{k-1}\alpha _{k-1},\lambda _k)$.\endproclaim

\demo{Proof} The free right action is given by $(\alpha _{k-1},\lambda _k)\cdot (\eta _{k-2},\sigma _{k-1}):=(\alpha _{k-1}+d_{k-2}\eta
_{k-2}+\sigma _{k-1},\lambda _k)$, where $(\alpha _{k-1},\lambda _k)\in\Cal G^k$ and $(\eta _{k-2},\sigma _{k-1})\in\Cal G_{\ast}^{k-1}$. With
respect to the open cover $\Cal V^{(k-1)}$ of $\Bbb T^{b_{k-1}}$ (3.9), the bundle atlas is

$$\split & \psi _a^{(k-1)} \colon V_a^{(k-1)}\times\Cal G_{\ast}^k\times\Cal G_{\ast}^{k-1}\rightarrow \tilde\pi _{k-1}^{-1}(V_a^{(k-1)}\times
\Cal G_{\ast}^k)\subseteq\Cal G^{k}\\ & \psi _a^{(k-1)}(z_1,\ldots ,z_{b_{k-1}},\xi _{k-1},\lambda _k,\eta _{k-2},\sigma _{k-1})=(\sum
_{j=1}^{b_{k-1}}s_{a_j}(z_j)\rho _j^{(k-1)}+\xi _{k-1}+d_{k-2}\eta _{k-2}+\sigma _{k-1},\lambda _k)
\\ &(\psi _a^{(k-1)})^{-1}(\alpha _{k-1},\lambda _k)=(\pi _{k-1}(\alpha _{k-1},\lambda _k),\omega _a^{(k-1)}(\alpha _{k-1})).
\endsplit\tag4.18$$ \qed\enddemo

One obtains for the induced volume form

$$(\psi _a^{(k-1)})^{\ast}vol_{\Cal G^k}=\frac{1}{(2\pi )^{b_{k-1}}}\det{(h^{(k-1)})}^{1/2}\det{(\Delta _{k-2}|_{imd_{k-1}^{\ast}})}^{1/2}vol_{\Bbb
T^{b_{k-1}}}\wedge vol_{\Cal G_{\ast}^k}\wedge vol_{\Cal G_{\ast}^{k-1}}.\tag4.19$$ Finally, $Q^{(k)}:=\sum\limits _{a\in\Bbb
Z_2^{b_{k-1}}}(pr_1\circ\pi _{k-1})^{\ast}g_a^{(k-1)}e^{-(pr_2\circ\pi _{k-1})^{\ast}S_{reg}^{(k)}-(\omega _a^{(k-1)})^{\ast}S_{reg}^{(k-1)}}$
regularizes the volume of $\Cal G^k$ for any $k=1,\ldots ,p$, where $pr_1$ and $pr_2$ denote the projections from $\Bbb T^{b_{k-1}}\times\Cal
G_{\ast}^{k}$ onto the first and the second factor, respectively. The volume of $\Cal G^k$, $k=1,\ldots ,p$, reads

$$Vol(\Cal G^k;Q^{(k)})=(\det{h^{(k-1)}})^{1/2}
(\det{\Delta _{k-2}|_{imd_{k-1}^{\ast}}})^{1/2} Vol(\Cal G_{\ast}^k;e^{-S_{reg}^{(k)}})Vol(\Cal G_{\ast}^{k-1};e^{-S_{reg}^{(k-1)}}).\tag4.20$$
Thus the volume of the restricted gauge group $\Cal G_{\ast}^k$ can be expressed in terms of the volume of the restricted gauge group $\Cal
G_{\ast}^{k-1}$ of one degree lower. By induction on (4.20) and using (4.17), one finds for $p\neq 0$

$$\multline\frac{Vol(\Cal G_{\ast}^p;e^{-S_{reg}^{(p)}})}{Vol(\Cal G^p;Q^{(p)})}=\left(Vol(kerd_{p-1},\tilde Q^{(p-1)}\right)^{-1}=
\\=\prod _{j=0}^{p-1} (\det h^{(j)})^{\frac{1}{2}(-1)^{p-j}}\prod _{j=0}^{p-2}(\det{\Delta _j|_{imd_{j+1}^{\ast}}})^{\frac{1}{2}(-1)^{p+1-j}}\prod
_{j=0}^{p-1}Vol(\Cal G^j;Q^{(j)})^{(-1)^{p-j}}.\endmultline\tag4.21$$ For $p=0$, one has $Vol(\Cal G_{\ast}^p;e^{-S_{reg}^{(p)}})=Vol(\Cal
G^p;Q^{(p)})$. \par

Eq. (4.21) can be interpreted as generalization of the ghost-for-ghost contribution, which has been derived in the topologically trivial
context based on either the Faddeev-Popov approach [12-15] or the technique of resolvents of differential operators [17]. Geometrically, this
contribution traces back to the non-free action of the gauge group $\Cal G^p$ on $\Cal A^p$. In our analysis, (4.21) is the result of
subsequent fiber integrations. In the following we choose the functionals $Q^{(j)}$ such that $Vol(\Cal G^j;Q^{(j)})=1$.

\subheading{Independence of the background connection} For the moment the expression $\Cal I^{(p)}(f)$ in (4.4) seems to depend on the choice
for the family of background differential characters. Let us take a different family $\hat v_0^c$ with $\delta _2(\hat v_0^c)=c$. Hence there
exists a family of classes $[B^c]\in\Cal M^p$ with $d_pB^c=0$ such that $\hat v_0^c=\hat u_0^c\cdot j_2([B^c])$. According to theorem 3.3 one
can find a $w:=(w_1,\ldots ,w_{b_p})\in\Bbb T^{b_p}$ such that $[B^c]=\chi _a(w_1,\ldots ,w_{b_p},0)$. If we define the left translation
$l_{w}$ on $\Bbb T^{b_p}$ by $l_{w}(z):=(w_1z_1,\ldots ,w_{b_p}z_{b_p})$, then $\tilde f^{\hat v_0^c}=l_{w}^{\ast}\tilde f^{\hat u_0^c}$ which
finally yields $\tilde f_{(0,\dots ,0)}^{\hat v_0^c}=\tilde f_{(0,\dots ,0)}^{\hat u_0^c}$. Together with $\tilde S_{inv}^{\hat v_0^c}=\tilde
S_{inv}^{\hat u_0^c}$ we get $\Xi _a(\hat v_0^c)=\Xi _a(\hat u_0^c)$ and so we have proved that $\Cal I^{(p)}(f)$ is indeed independent of the
chosen set of background differential characters.

\subheading {Sum over topological sectors} Since the cohomology of $M$ is finitely generated, $c\in H^{p+1}(M;\Bbb Z)$ admits the following
(non-canonical) decomposition

$$c=\sum _{j=1}^{b_{p+1}}m_jf_{j}^{(p+1)}+
\sum _{k=1}^{r}y_kt_{k}^{(p+1)},\tag4.22$$ where $(f_{j}^{(p+1)})_{j=1}^{b_{p+1}}$ denotes a Betti basis of $H^{p+1}(M;\Bbb Z)$ and $m_j\in\Bbb
Z$. Furthermore $TorH^{p+1}(M;\Bbb Z)$ is generated by a basis $(t_{k}^{(p+1)})_{k=1}^{r}$ and $y_k\in\Bbb Z_{l_k}$. By definition there exists
a series of elements $l_1,\ldots ,l_r\in\Bbb N$ such that $l_kt_{k}^{(p+1)}=0$ for each $k=1,\ldots ,r$. Evidently, the order of the torsion
subgroup, denoted by $ord(TorH^{p+1}(M;\Bbb Z))$, is given by $\prod _{k=1}^rl_k$. Let $\rho _j^{(p+1)}\in Harm_{\Bbb Z}^{p+1}(M;\Bbb R)$, for
$j=1,\ldots ,b_{p+1}$, be a basis of harmonic $(p+1)$-forms on $M$ with integer periods, and let $h_{jk}^{(p+1)}=<\rho _j^{(p+1)},\rho
_k^{(p+1)}>$ denote the induced metric on $Harm^{p+1}(M;\Bbb R)$. By lemma 4.1 one has $\delta _1(\hat u_0^c)\in Harm_{\Bbb Z}^{p+1}(M;\Bbb
R)$, so that

$$\delta
_1(\hat u_0^c)=\sum _{k=1}^{b_{p+1}}m_k\rho _k^{(p+1)},\quad m_k=\sum _{j=1}^{b_{p+1}}(h_{jk}^{(p+1)})^{-1}<\delta _1(\hat u_0^c),\rho
_{j}^{(p+1)}> \in\Bbb Z.\tag4.23$$ According to the Hodge decomposition theorem and the fact that $\det{\Delta _{p+1}|_{imd_{p}}}=\det{\Delta
_p|_{imd_{p+1}^{\ast}}}$ [25], the determinant of the restricted Laplacian $\Delta _p|_{Harm^p(M)^{\perp}}$ factorizes into

$$\det{(\Delta _p|_{Harm^p(M)^{\perp}})}=\det{(\Delta _p|_{imd_{p+1}^{\ast}})}\cdot\det{(\Delta _{p-1}|_{imd_{p}^{\ast}})}.
\tag4.24$$ By induction one obtains

$$\prod _{j=0}^p(\det{\Delta _j|_{imd_{j+1}^{\ast}}})^{\frac{1}{2}(-1)^{p+1-j}}=
\prod _{j=0}^p(\det{\Delta _j|_{Harm^j(M)^{\perp}}})^{\frac{1}{2}(-1)^{p+1-j}(p+1-j)}.\tag4.25$$ Till now there is one step left, namely the
summation over the cohomology classes in $H^{p+1}(M;\Bbb Z)$. According to (4.22) this sum is split into two parts, one over the components of
the free part, the other one over the torsion part of $c$. In order to perform this sum, let us recall the definition of the Riemann Theta
function: For $\Lambda$ being a symmetric complex $k\times k$ dimensional square matrix whose imaginary part is positive definite, $b\in\Bbb
C^k$ the $k$-dimensional Theta function is defined by

$$\Theta _{k}(b|\Lambda )=\sum\limits _{n\in\Bbb Z^k}\exp{\lbrace \pi\sqrt{-1}n
^{\dag}\cdot\Lambda\cdot n+2\pi\sqrt{-1}n^{\dag}\cdot b\rbrace },\tag4.26$$ where the superscript $\dag$ denotes the transpose. Evidently, the
Theta function possesses the symmetry $\Theta _{k}(b+m|\Lambda )=\Theta _{k}(b|\Lambda )$ for all $m\in\Bbb Z^k$.

\proclaim{Theorem 4.5} For the generalized $p$-form Maxwell theory, the partition function $\Cal Z^{(p)}$, $0\leq p\leq n$,  is given by
$$\Cal Z^{(p)}=\prod _{j=0}^p
\left(\frac{(\det{\Delta _j|_{Harm^j(M)^{\perp}}})^{(p+1-j)}}{\det h^{(j)}} \right)^{\frac{1}{2}(-1)^{p+1-j}}\Theta _{b_{p+1}}
\left(0|-\frac{h^{(p+1)}}{2\pi\sqrt{-1}}\right) ord(TorH^{p+1}(M;\Bbb Z)).\tag4.27$$ The VEV of a gauge invariant observable $f$ admits the
following form: \roster\item For $p=0$:
$$\Cal E^{(0)}(f)=\frac{(\det{\Delta _0|_{imd_{1}^{\ast}}})^{1/2}
\sum\limits _{c\in H^{1}(M;\Bbb Z)}e^{-\frac{1}{2}||(\hat u_0^c)^{\ast}\vartheta ||^2}\ \int\limits _{imd_{1}^{\ast}} \Cal D\tau _{0}\
\tilde{f}^{\hat u_0^c}_{(0)}(\tau _0)\cdot e^{-\frac{1}{2}<\tau _0,\Delta _0|_{imd_{1}^{\ast}}\tau _0>}}{\Theta _{b_{1}}
\left(0|-\frac{h^{(1)}}{2\pi\sqrt{-1}}\right)} \tag4.28a$$\item For $p\neq 0$:

$$\multline\Cal E^{(p)}(f)=\\ =\frac{(\det{\Delta _p|_{Harm^p(M)^{\perp}}})^{1/2} \sum\limits _{c\in H^{p+1}(M;\Bbb Z)}e^{-\frac{1}{2}||\delta _1(\hat
u_0^c)||^2}\ \int\limits _{imd_{p+1}^{\ast}} \Cal D\tau _{p}\ \tilde{f}^{\hat u_0^c}_{(0,\ldots ,0)}(\tau _p)\cdot e^{-\frac{1}{2}<\tau
_p,\Delta _p|_{imd_{p+1}^{\ast}}\tau _p>}}{(\det{\Delta _{p-1}|_{imd_{p}^{\ast}}})^{1/2}\ \Theta _{b_{p+1}}
\left(0|-\frac{h^{(p+1)}}{2\pi\sqrt{-1}}\right)\ ord(TorH^{p+1}(M;\Bbb Z))}.\endmultline\tag4.28b$$\endroster\qed\endproclaim

Due to the invariance under unimodular transformations the partition function is independent of the chosen basis of harmonic forms. \par

For manifolds which are acyclic for all j satisfying $0<j\leq p+1<n$, one can choose $\hat{u}_0^c=1$. It follows from theorem 3.3 that $\Cal
M^{p}\cong imd_{p+1}^{\ast}$, implying that the gauge orbit space coincides with the space of transversal fields. Then (4.27) reduces to

$$\Cal Z^{(p)} =Vol(M)^{\frac{1}{2}(-1)^p}\prod _{j=0}^p
\left(\det{\Delta _j|_{imd_{j+1}^{\ast}}}\right)^{\frac{1}{2}(-1)^{p+1-j}},\tag4.29$$ which up to the finite volume of $M$ agrees with the
known result for the partition function of the $p$-form Maxwell theory [18]. The reason is that the gauge groups are $imd_{j-1}\times
Harm_{\Bbb Z}^j(M)$ in our setting compared with $imd_{j-1}$ used in [15,18] for $j=0,\ldots ,p$. These groups agree for $j\neq 0$ in the
acyclic case.\par

Additionally, we recover the duality relation between the partition functions in various degrees: Let us begin with the definition of the
Ray-Singer analytic torsion $\tau (M)$ of the Riemannian manifold $M$ [26]

$$\tau (M):=\exp{\left( \frac{1}{2}\sum _{k=0}^n(-1)^{k}k\frac{d}{ds}|_{s=0}\zeta (s|\Delta _k)\right)}=\prod _{k=0}^{n}
(\det{\Delta _{k}|_{Harm^k(M)^{\perp}}})^{(-1)^{k+1}\frac{k}{2}},\tag4.30$$ which by (4.24) can be rewritten into the product $\tau (M)=\prod
_{k=0}^{n-1}(\det{\Delta _k|_{imd_{k+1}^{\ast}}})^{\frac{1}{2}(-1)^{k}}$. Using the fact that $\star\Delta _{p}=\Delta _{n-p}\star$ holds, one
can verify easily that

$$\det{\Delta _{j}|_{imd_{j+1}^{\ast}}}=\det{\Delta _{n-j-1}|_{imd_{n-j}^{\ast}}}.\tag4.31$$ From (4.29) we finally obtain the following
duality relation [18],

$$\frac{\Cal Z^{(p)}}{\Cal Z^{(n-p-2)}}=\left(\frac{\tau (M)}{Vol(M)^{\frac{1}{2}(1-(-1)^n)}}\right)^{(-1)^{p+1}}.\tag4.32$$ In even
dimension $\tau (M)=1$ and thus the partition functions $\Cal Z^{(p)}$ and $\Cal Z^{(n-p-2)}$ coincide.\par

In summary, a modified functional integral has been introduced for the quantization of generalized abelian gauge theories. It was then shown
that the VEV of gauge invariant observables \roster\item is independent of the choice for the regularization of the gauge groups, \item is
independent of the local trivialization, \item is independent of the choice for the partition of unity, \item reproduces the conventional
result for acyclic manifolds, yet the volume of the gauge group can be absorbed into a finite normalization constant.\endroster

\subheading {An explicit choice for the regularizing functions} The results (4.27) and (4.28) rely on the assumption that appropriate
regularizing functions $S_{reg}^{(k)}$ and $Q^{(k)}$ do really exist. Let us define:\pagebreak\roster\item For k=0:

$$e^{-S_{reg}^{(0)}(\lambda _0)}=Q^{(0)}(\lambda _0)=\Theta _{1} \left( 0|-\frac{Vol(M)}{2\pi\sqrt{-1}}\right)^{-1}
e^{-\frac{1}{2}\lambda _0^2Vol(M)},\quad\lambda _0\in\Bbb Z.\tag4.33$$\item For $k=1,\ldots ,p$:

$$e^{-S_{reg}^{(k)}(\xi _{k-1},\lambda _{k})}=\prod _{l=0}^{k-1}\left(\frac{\det{\Delta
_{l}|_{imd_{l+1}^{\ast}}}}{\det{h^{(l)}}}\right)^{\frac{1}{2}(-1)^{k+1-l}}\frac{\det{(\Delta _{k-1}|_{imd_{k}^{\ast}})}^{1/2}\
e^{-\frac{1}{2}\|\Delta _{k-1}|_{imd_{k}^{\ast}}\xi _{k-1}\|^2-\frac{1}{2}\|\lambda _{k}\|^2}}{\Theta _{b_{k}} \left(
0|-\frac{h^{(k)}}{2\pi\sqrt{-1}}\right)}.\tag4.34$$\endroster By a direct calculation one finds that

$$\sum _{\lambda _k\in Harm_{\Bbb Z}^k(M)}\int _{imd_{k}^{\ast}}\Cal D\xi _{k-1}
e^{-\frac{1}{2}\|\Delta _{k-1}|_{imd_{k}^{\ast}}\xi _{k-1}\|^2-\frac{1}{2}\|\lambda _{k}\|^2}=(\det{\Delta _{k-1}|_{imd_k^{\ast}}})^{-1}\Theta
_{b_k} \left(0|-\frac{h^{(k)}}{2\pi\sqrt{-1}} \right)\tag4.35$$ which in summary leads to the following finite volumes of the gauge groups

$$Vol(\Cal G_{\ast}^{k};e^{-S_{reg}^{(k)}})=\cases 1 &\text{if $k=0$} \\(\det{(\Delta _{k-1}|_{imd_{k}^{\ast}})}^{-\frac{1}{2}}
\ \prod _{l=0}^{k-1}\left(\frac{\det{\Delta _{l}|_{imd_{l+1}^{\ast}}}}{\det{h^{(l)}}}\right)^{\frac{1}{2}(-1)^{k+1-l}} &\text{if $k\neq
0$}.\endcases\tag4.36$$ Eq. (4.20) implies that $Vol(\Cal G^{(k)},Q^{(k)})=1$ for all $k=0,\ldots p$. Given this choice, we are able to write
down the partition function (4.5) in the original field space $\Cal A^p$. Using (3.13), one finds

$$\Cal Z^{(p)}=\int _{\Cal A^p}vol_{\Cal A^p}\ \Cal F^{(p)}(A)\ e^{-\frac{1}{2}<A,\Delta _pA>},\tag4.37$$ with a non-negative functional
$\Cal F^{(p)}(A)$, which for $p=0$ yields

$$\Cal F^{(0)}(A)=\frac{\Theta _{b_{1}} \left(0|-\frac{h^{(1)}}{2\pi\sqrt{-1}}\right)}
{\Theta _{1} \left(0|-\frac{Vol(M)}{2\pi\sqrt{-1}}\right)}\sum _{a_1=1}^2\tilde g_{a_1}(e^{2\pi\sqrt{-1}\int _{M}A\wedge\rho ^{(n)}})
e^{-\frac{1}{2}Vol(M)(\epsilon _{a_1}^{(0)}(A))^2},\tag4.38$$ and for $p\neq 0$ this functional is given by

$$\split\Cal F^{(p)}(A) &=\frac{\Theta
_{b_{p+1}} \left(0|-\frac{h^{(p+1)}}{2\pi\sqrt{-1}}\right)}{\Theta _{b_{p}} \left(0|-\frac{h^{(p)}}{2\pi\sqrt{-1}}\right)}(\det{\Delta
_{p-1}|_{imd_{p}^{\ast}}})^{1/2}\prod _{j=0}^{p-1}\left(\frac{\det{\Delta _j|_{imd_{j+1}^{\ast}}}}{\det h^{(j)}}
\right)^{\frac{1}{2}(-1)^{p+1-j}}\\ &\times\sum _{a_1=1}^2\ldots\sum _{a_{b_p}=1}^2\tilde g_{a_1}(e^{2\pi\sqrt{-1}\int _{M}A\wedge\rho
_1^{(n-p)}})\cdots\tilde g_{a_{b_p}}(e^{2\pi\sqrt{-1}\int _{M}A\wedge\rho _{b_p}^{(n-p)}}) e^{-\frac{1}{2}\sum
_{j,k=1}^{b_p}h_{jk}^{(p)}\epsilon _{a_j}^{(p)}(A)\epsilon _{a_k}^{(p)}(A)}\\ &\times ord(TorH^{p+1}(M;\Bbb Z)).\endsplit\tag4.39$$

If $M$ is acyclic in dimension $0<k\leq p+1<n$, the bundle $\Cal A^p\rightarrow\Cal A^p/imd_{p-1}$ is trivializable and the bundle chart (3.14)
gives the Hodge decomposition of $A\in\Cal A^p$. This guarantees the existence of a global smooth gauge fixing submanifold in $\Cal A^p$. The
corresponding partition function is

$$\Cal Z^{(p)}=(Vol(M))^{\frac{1}{2}(-1)^p}(\det{\Delta _{p-1}|_{imd_p^{\ast}}})\prod _{j=0}^{p-2}
(\det{\Delta _{j}|_{imd_{j+1}^{\ast}}})^{\frac{1}{2}(-1)^{p+1-j}}\int _{\Cal A^p}vol_{\Cal A^p}e^{-\frac{1}{2}<A,\Delta _pA>}.\tag4.40$$ For
$p=0$, (4.37) reduces simply to the partition function for (real-valued) scalar fields. Up to the factor $Vol(M)$, (4.40) compares to [15],
where the partition function for the $p$-form Maxwell theory has been derived based on the Faddeev-Popov technique in the original field space
$\Cal A^p$.\par

In the topologically non-trivial case, the functional $\Cal F^p(A)$ guarantees finiteness of (4.37). In fact, if one considered the
conventional gauge fixing term $\frac{1}{2}\|d_{p}^{\ast}A\|^2$ coming from the Faddeev-Popov approach, then the functional integral $\int
_{\Cal A^p} \exp{(-\frac{1}{2}<A,\Delta _pA>)}$ would become infinite. This can be easily seen by rewriting this integral in terms of the local
trivialization of $\Cal A^p@>>>\Cal M^p$ and is a consequence of the fact that the integrand is not damped along gauge transformations not
connected to unity. Our method solves this problem by introducing an appropriate regularization, however, without affecting the VEV of gauge
invariant observables.
\bigskip
{\bf 5. The Green\rq s functions for the generalized $p$-form Maxwell theory}
\bigskip
In this section we want to determine the one-point- and two-point functions for the gauge field $A\in\Cal A^p$ in the generalized $p$-form
Maxwell theory. Since these functions are not gauge invariant, one could expect additional contributions resulting from the non-trivial
structure of configuration space and the regularization of the gauge group.\par

Like in the $p$-form Maxwell theory, the Green\rq s functions are generated by the the vacuum-to-vacuum transition amplitude in the presence of
a source $J\in \Omega ^p(M;\Bbb R)$ which in our approach takes the form

$$\Cal Z^{(p)}[J]=\frac{1}{Vol(\Cal G^p;Q^{(p)})}\sum\limits _{c\in H^{p+1}(M;\Bbb Z)}\int _{\Cal A^{p}}\sum _{a\in \Bbb Z_2^{b_p}}
((\pi _{\Cal A^p})^{\ast}\hat g_a^{(p)})\ \Xi _a(\hat{u}_0^c,A)\ e^{<J,A>}.\tag5.1$$ The $q$-point Green\rq s functions $\Cal S_q^{(p)}$ are
defined by

$$\Cal S_q^{(p)}(v_1\ldots ,v_q) :=\frac{\partial ^q}{\partial t_1\cdots\partial t_q}
\vert _{t_1=\ldots =t_q=0} \ \frac{\Cal Z^{(p)}[\sum _{i=1}^qt_iv_i]}{Z^{(p)}[0]},\tag5.2$$ for $v_1,\ldots ,v_q\in \Omega ^p(M;\Bbb R)$. By
construction, the Green\rq s functions are independent of the chosen background differential character. In order to derive an explicit
expression for (5.1), we take the choice (4.30) and (4.31) for $S_{reg}^{(k)}$, $k=0,\ldots ,p$. In terms of the local trivialization $\{\psi
_a\}$ a lengthy calculation gives

$$\split \Cal Z^{(p)}[J]&=\frac{1}{(2\pi )^{b_p}}\prod _{j=0}^p \left(\frac{\det{\Delta _j|_{imd_{j+1}^{\ast}}}}{\det h^{(j)}}
\right)^{\frac{1}{2}(-1)^{p-1-j}}\cdot e^{-\frac{1}{2}<J,G_pJ>}\\ &\times \frac{\Theta _{b_{p}}
\left(K^{(p)}(J)|-\frac{h^{(p)}}{2\pi\sqrt{-1}}\right)\Theta _{b_{p+1}} \left(0|-\frac{h^{(p+1)}}{2\pi\sqrt{-1}}\right)}{\Theta _{b_{p}}
\left(0|-\frac{h^{(p)}}{2\pi\sqrt{-1}}\right)}\cdot ord(TorH^{p+1}(M;\Bbb Z))\\ & \times \int _{\Bbb T^{b_p}}vol_{\Bbb T^{b_p}} \sum
_{a_1=1}^2\cdots\sum _{a_{b_p}=1}^2 q_{1}^{\ast}\tilde g_{a_1}\cdots q_{b_p}^{\ast}\tilde g_{a_{b_p}}\cdot e^{2\pi\sum
_{j=1}^{b_p}q_{j}^{\ast}s_{a_{j}}<J,\rho _j^{(p)}>},\endsplit\tag5.3$$ where $K_j^{(p)}(J):=<J,\rho _j^{(p)}>$ with $j=1,\ldots ,b_p$ is
regarded as $b_p$-dimensional vector, denoted by $K^{(p)}(J)$. Defining the two field independent factors

$$\split &\varepsilon _{j}^{(1)}:=\frac{1}{2\pi}\int _{\Bbb T^{1}} vol_{\Bbb T^{1}}
\sum _{a_{j}=1}^2\tilde g_{a_{j}} s_{a_{j}} \\
&\varepsilon _{j,k}^{(2)}:= \frac{1}{(2\pi)^{b_p}}\int _{\Bbb T^{b_p}} vol_{\Bbb T^{b_p}} \sum _{a_{1}=1}^2\ldots\sum _{a_{b_p}=1}^2
q_{1}^{\ast}\tilde g_{a_{1}}\cdots q_{b_p}^{\ast}\tilde g_{a_{b_p}} \cdot q_{j}^{\ast}s_{a_{j}}\cdot q_{k}^{\ast}s_{a_{k}}\endsplit\tag5.4$$ one
finally ends up with the following result:

\proclaim{Proposition 5.1} Let us regularize the gauge group by (4.33) and (4.34). Then the following holds for the Green\rq s functions:
\par 1) One-point function:

$$\Cal S_1^{(p)}(v) =\sum
_{j=1}^{b_p}\varepsilon _{j}^{(1)} <v,\rho _j^{(p)}>+ \frac{d}{dt}|_{t=0}\ln{\Theta
_{b_p}(K^{(p)}(tv)|-\frac{h^{(p)}}{2\pi\sqrt{-1}})}.\tag5.5$$

2) Two-point function:

$$\split \Cal S_2^{(p)}(v_1,v_2)= &<v_1,
G_pv_2>+\sum _{j=1}^{b_p}\varepsilon _{j}^{(1)}<v_{1},\rho _j^{(p)}>\frac{d}{dt}|_{t=0}\ln{\Theta _{b_p}\left( K^{(p)}(tv_2
)|-\frac{h^{(p)}}{2\pi\sqrt{-1}} \right)})\\ &+ \sum _{j=1}^{b_p}\varepsilon _{j}^{(1)}<v _{2},\rho _j^{(p)}>\frac{d}{dt}|_{t=0}\ln{\Theta
_{b_p}\left( K^{(p)}(tv_1 )|-\frac{h^{(p)}}{2\pi\sqrt{-1}} \right)}) \\ &+ \sum _{j,k=1}^{b_p}\varepsilon _{j,k}^{(2)}<v_{1},\rho _j^{(p)}><v
_{2},\rho _k^{(p)}> \\ &+\Theta _{b_p}\left( 0|-\frac{h^{(p)}}{2\pi\sqrt{-1}}\right)^{-1}\frac{\partial ^2}{\partial t_1\partial t_2} \vert
_{t_1=t_2=0}\Theta _{b_p}\left( K^{(p)}(\sum _{l=1}^2t_lv_l)|-\frac{h^{(p)}}{2\pi\sqrt{-1}} \right)\endsplit\tag5.6$$ \qed\endproclaim

On manifolds with vanishing $p$-th Betti number, the one-point function vanishes, whereas the two-point function reduces to the Green\rq s
operator $G_p$.\par The numerical factors in (5.4) can be easily calculated in terms of a natural partition of unity for $\Bbb T^1$. Taking the
following local coordinate system of $\Bbb T^1$
$$\align v_{1}\colon V_{1} \rightarrow (0,1 )\quad v_{1}^{-1}(t) &
=(\cos{2\pi t},\sin{2\pi t})\\  v_{2}\colon V_{2} \rightarrow (-\frac{1}{2} ,\frac{1}{2} )\quad v_{2}^{-1}(t) & =(\cos{2\pi t},\sin{2\pi
t}).\tag5.7\endalign$$ A partition of unity subordinate to $V_{j}\subset\Bbb T^1$ is induced by the periodic functions $\hat g_1(t)=\sin ^2(\pi
t)$ and $\hat g_2(t)=\cos ^2(\pi t)$. A calculation yields for (5.4)

$$\split\varepsilon _{j}^{(1)} &=\frac{3}{4}\\
\varepsilon _{j,k}^{(2)} &= \cases (2\pi )^{-1}(\frac{17\pi}{12}-\frac{1}{\pi}), &\text{for $j=k$ }\\ (2\pi )^{-2}(\frac{3\pi}{2})^2, &\text{for
$j\neq k$ }.\endcases
\endsplit\tag5.8$$ How do these factors depend on the choice of the local section $s_a$ in (3.10)? Any other local section $s_a^{\prime}$ of
(3.8) is connected with the section $s_a$ by $s_a^{\prime}:=(s_{a_{1}}^{\prime},\ldots ,s_{a_{b_p}}^{\prime})=(s_{a_{1}}+m_{a_{1}}^{1},\ldots
,s_{a_{b_p}}+m_{a_{b_1}}^{b_p})$ with $m_{a_{j}}^{j}\in\Bbb Z$ for $j=1,\ldots b_p$. In terms of these new sections, the factors (5.4) read

$$\split\varepsilon
_{j}^{\prime (1)} &=\frac{1}{2} (m_{1}^{j}+ m_{2}^{j}+\frac{3}{2}) \\
\varepsilon _{j,k}^{\prime (2)} &= \cases (2\pi )^{-1}\left(\frac{17\pi}{12}-\frac{1}{\pi}+\pi
(m_{1}^{j}(m_{1}^{j}+1)+m_{2}^{j}(m_{2}^{j}+2))\right),\quad\text{for $j=k$}
\\ \frac{1}{4}(m_{1}^{j}+m_{2}^{j}+\frac{3}{2})(m_1^{k}+m_{2}^{k}+\frac{3}{2}),\quad\text{for $j\neq k$ }.\endcases\endsplit\tag5.9$$ Thus it is not
possible to arrange a local trivialization of $\Cal A^{p}$ in such a way that the topological contributions in the Green\rq s functions would
vanish. The non-vanishing of the one-point function and the occurrence of additional contributions in the two-point function are caused by the
non-trivial geometric structure of the configuration space and the finiteness of the volume of the gauge degrees of freedom.

\subheading{One-point functions in special dimensions} Let us conclude with two simple examples for the one-point function for $p=0$ and $p=n$,
respectively. Since $\rho ^{(0)}=1$ and $\rho ^{(n)}=\frac{vol_M}{Vol(M)}$, the corresponding one-point functions read
$$\split &\Cal S_1^{(0)}(w) = \frac{3}{4}\int _M\star w,\qquad w\in\Omega ^0(M;\Bbb R),\\
&\Cal S_1^{(n)}(v) = \frac{3}{4}\frac{1}{Vol(M)}\int _Mv,\qquad v\in\Omega ^n(M;\Bbb R).\endsplit\tag5.10$$
\bigskip\bigskip
{\bf 6. The VEV of special gauge invariant observables}
\bigskip
{\bf 6.1. Smooth homomorphisms and the Poincar\'e-Pontrjagin duality}\bigskip
\bigskip
The Poincar\'e-Pontrjagin duality for differential characters induces a specific class of gauge invariant observables in a natural way. In
fact, each $\hat v\in\hat H^{n-p-1}(M;\Bbb R/\Bbb Z)$ gives rise to a homomorphism $\Cal D(\hat v)\in\hat H^p(M;\Bbb R/\Bbb Z)^{\ast}$. Now we
will study the VEV of these gauge invariant observables.\par

By theorem 2.2 we choose a set of differential characters for $\hat f_j^{(p)}$, with $j=1,\ldots ,b_{p+1}$, and $\hat t_k^{\prime (p)}$, with
$k=1,\ldots ,r$, in $\hat H^{p}(M;\Bbb R/\Bbb Z)$ such that

$$\left. \aligned \delta _1(\hat f_j^{(p)}) &=\rho _j^{(p+1)}\\\delta _1(\hat t_k^{\prime (p)}) &=0\endaligned\qquad\qquad\aligned
\delta _2(\hat f_j^{(p)})&=f_j^{(p+1)} \\ \delta _2(\hat t_k^{\prime (p)})& =t_k^{(p+1)}.\endaligned\right.\tag6.1$$ The last line of (6.1)
implies that there exists $[v_k^{\prime}]\in H^p(M;\Bbb R/\Bbb Z)$ such that $\hat t_k^{\prime (p)}=j_1([v_k^{\prime}])$ for $k=1,\ldots ,r$.
Moreover, since $\delta _2((\hat t_k^{\prime (p)})^{l_k})=0$ with torsion coefficients $l_k$, there exists a family of closed differential
forms $B_k\in\Omega ^p(M;\Bbb R)$ such that $(\hat t_k^{\prime (p)})^{l_k}=j_2([B_k])$. Hence $\hat t_k^{(p)}:=\hat t_k^{\prime (p)}\cdot
j_2([-\frac{1}{l_k}B_k])$ fulfills (6.1) but satisfies $(\hat t_k^{(p)})^{l_k}=1$ for each $k=1,\ldots ,r$. Hence there are cohomology classes
$[v_k]\in H^{p}(M;\Bbb R/\Bbb Z)$ such that $\hat t_k^{(p)}=j_1([v_k])$. In summary we will take the following choice for the background
differential character

$$\hat u_0^c=\prod _{j=1}^{b_{p+1}}(\hat f_j^{(p)})^{m_j}\prod _{k=1}^{r}(\hat t_k^{(p)})^{y_k}.\tag6.2$$  Coming back to (4.28) the zeroth
Fourier coefficient of $\Cal D(\hat v)$ is given by

$$\split\tilde \Cal D(\hat v)_{(0,\ldots ,0)}^{\hat u_0^c}(\tau _p) &=\Cal D(\hat v)(\hat u_0^c)e^{2\pi\sqrt{-1}(-1)^{n-p}
\int _{M}\delta _1(\hat v)\wedge\tau _p} \prod _{j=1}^{b_p}\int _{0}^1dw_j\ e^{2\pi\sqrt{-1}(-1)^{n-p}w_j\int _{M}\delta _1(\hat v)\wedge
\rho _j^{(p)}} \\
&=\cases 0 &\text{if $\int _{M}\delta _1(\hat v)\wedge\rho _j^{(p)}\neq 0$}\\ \Cal D(\hat v)(\hat u_0^c)e^{2\pi\sqrt{-1}(-1)^{n-p}\int
_{M}\delta _1(\hat v)\wedge\tau _p} &\text{if $\int _{M}\delta _1(\hat v)\wedge\rho _j^{(p)}=0$,}\endcases\endsplit\tag6.3$$ where $j=1,\ldots
,b_p$. In order to get a non-vanishing $\Cal I^{(p)}(\Cal D(\hat v))$ one has to demand that $[\delta _1(\hat v)]=0$ in $H^{n-p}(M;\Bbb R)$,
implying that $\delta _2(\hat v)\in\ker r_{\ast}$, where the third exact sequence of theorem 2.2 has been used. According to the exact sequence
in (2.3) there exists a class $[w]\in H^{n-p-1}(M;\Bbb R/\Bbb Z)$, such that $\delta ^{\ast}([w])=\delta _2(\hat v)$. From the exact sequences
in theorem 2.2 we can finally conclude that there exists a $B\in\Omega ^{n-p-1}(M;\Bbb R)$ such that $\hat v=j_1([w]^{-1})j_2([B])$. Inserting
this expression for $\hat v$ and using the properties of the product in (2.4), we find after a straightforward calculation

$$\Cal D(\hat v)(\hat u_0^c)=<j_1([w]^{-1}\cup c),[M]>\ e^{2\pi\sqrt{-1}\int _MB\wedge\delta _1(\hat u_0^c)}.\tag6.4$$ Together with
the decomposition (4.22) and (4.23) we are now ready to perform the summation over the cohomology classes $c\in H^{p+1}(M;\Bbb Z)$ in (4.28),
namely

$$\split\sum _{c\in H^{p+1}(M,\Bbb Z)}\Cal D(\hat v)(\hat u_0^c)e^{-\frac{1}{2}\|\delta _1(\hat u_0^c)\|^2}= &\sum _{m_1\in\Bbb Z}\cdots
\sum _{m_{b_{p+1}}\in\Bbb Z} \prod _{j=1}^{b_{p+1}}<[w]^{-1}\cup f_j^{(p+1)},[M]>^{m_j}\\ &\times e^{-\frac{1}{2}\sum
_{i,j=1}^{b_{p+1}}h_{ij}^{(p+1)}m_{i}m_{j}+2\pi\sqrt{-1}\sum _{j=1}^{b_{p+1}} m_j\int _MB\wedge\rho _j^{(p+1)}}\\
&\times \sum _{y_1=1}^{l_1-1}\cdots\sum _{y_r=1}^{l_r-1}\prod _{k=1}^{r}<[w]^{-1}\cup t_k^{(p+1)},[M]>^{y_k}\endsplit\tag6.5$$ where

$$\sum _{y_1=1}^{l_1-1}\cdots\sum _{y_r=1}^{l_r-1}\prod _{k=1}^{r}<[w]^{-1}\cup t_k^{(p+1)},[M]>^{y_k}=
\cases 0,&\text{if $<[w]^{-1}\cup t_k^{(p+1)},[M]>\neq 1$}\\ ord(TorH^{p+1}(M;\Bbb Z)),&\text{if $<[w]^{-1}\cup t_k^{(p+1)},[M]>=1$,
}\endcases\tag6.6$$ where $k=1,\ldots ,r$. In order to get a non-vanishing VEV, one has to demand that

$$<[w]^{-1}\cup t_k^{(p+1)},[M]>=<[w]^{-1},D(t_k^{(p+1)})>=1,\tag6.7$$ implying that $[w]$ vanishes on $TorH_{n-p-1}(M;\Bbb Z)$. Let us
define the map $\iota([w])\in Hom(H_p(M;\Bbb Z),\Bbb R/\Bbb Z)$ by $\iota([w])(\sigma ):=<[w],\sigma >$ for all $\sigma\in H_{n-p-1}(M;\Bbb
Z)$. Since $\iota([w])$ vanishes on torsion classes and $H_{n-p-1}(M;\Bbb Z)/TorH_{n-p-1}(M;\Bbb Z)$ is a free abelian group, it can be
extended to a homomorphism $\tilde{\iota}\in Hom(H_{n-p-1}(M;\Bbb Z),\Bbb R)$. By the universal coefficient theorem [24] there exist a $[\beta
]\in H^{n-p-1}(M;\Bbb R)$ such that $\tilde{\iota}(\sigma )=<[\beta ],\sigma >$. But then $[q\circ\beta ]$ gives a cohomology class in
$H^{n-p-1}(M;\Bbb R/\Bbb Z)$ which satisfies $\iota([q\circ\beta ])(\sigma )=<[w],\sigma >$ for all $\sigma\in H_{n-p-1}(M;\Bbb Z)$. Since
$\iota$ is an isomorphism between $H^{n-p-1}(M;\Bbb R/\Bbb Z)$ and $Hom(H_{n-p-1}(M;\Bbb Z),\Bbb R/\Bbb Z)$ it follows that $q_{\ast}([\beta
])=[q\circ\beta ]=[w]$. Let us represent the cohomology class $[\beta ]$ by a closed differential form $\beta\in\Omega ^{n-p-1}(M;\Bbb R)$,
then we get $\hat v=j_1(q_{\ast}([-\beta ]))j_2([B])=j_2([B-\beta ]$. In summary, we have verified:

\proclaim{Proposition 6.1} The VEV of the Poincar\'e dual differential character $\Cal D(\hat v)\in \hat H^{p}(M;\Bbb R/\Bbb Z)^{\ast}$
vanishes, unless $\hat v\in imj_2$.\qed\endproclaim

If we represent the real cohomology classes $f_j^{(p+1)}$ by harmonic forms, then the first factor on the right hand side of (6.5) becomes

$$<[w]^{-1}\cup f_j^{(p+1)},[M]>=q(<[-\beta]\cup f_j^{(p+1)},[M]>)=q(-\int _M\beta\wedge\rho _j^{(p+1)}).\tag6.8$$ Let us write
$C:=B-\beta$, then $\hat v=j_2([C])$ and $\delta _1(\hat v)=d_{n-p-1}C$. Inserting all our findings into (4.28) and performing the Gaussian
integral over $\tau _p\in imd_{p+1}^{\ast}$ finally gives

$$\Cal E^{(p)}(\Cal D(\hat v))=\frac{\Theta _{b_{p+1}} \left(K^{(p+1)}(C)|-\frac{h^{(p+1)}}{2\pi\sqrt{-1}}\right)}
{\Theta _{b_{p+1}} \left(0|-\frac{h^{(p+1)}}{2\pi\sqrt{-1}}\right)}\cdot e^{-\frac{(2\pi )^2}{2}<\star\delta _1(\hat v),G_{p}\star \delta
_1(\hat v)>},\tag6.9$$ where $K_j^{(p+1)}(C):=\int _{M}C\wedge\rho _j^{(p+1)}$ with $j=1,\ldots b_{p+1}$ is regarded as $b_{p+1}$-dimensional
vector. The choice of $C$ is unique up to elements in $\Omega _{\Bbb Z}^{n-p-1}(M;\Bbb R)$. Due to the invariance property of the Theta
function, any different choice for $C$ would give the same result for $\Cal E^{(p)}(\Cal D(\hat v))$.
\bigskip\bigskip

{\bf 6.2. The Wilson operator}\bigskip

The Wilson operator is a prominent example for a gauge invariant observable. For any $p$-cycle $\Sigma\in Z_p(M;\Bbb Z)$ with induced homology
class $[\Sigma ]\in H_p(M;\Bbb Z)$ the Wilson operator is defined by $W\colon Z_p(M;\Bbb Z)\rightarrow\hat H^p(M;\Bbb R/\Bbb Z)^{\ast}$,
$W(\Sigma )(\hat u):=\hat u(\Sigma )$. In contrast to the previous subsection, $W(\Sigma )$ is not in the smooth dual of $\hat H^p(M;\Bbb
R/\Bbb Z)$, i.e. does not belong to the range of the Poincar\'e-Pontrjagin map $\Cal D$ [22]. We begin with the calculation of the zeroth
Fourier coefficient of $W(\Sigma )$

$$\split\tilde W(\Sigma )_{(0,\ldots ,0)}^{\hat u_0^c}(\tau _p) &=\hat u_0^c(\Sigma )e^{2\pi\sqrt{-1}\int _{\Sigma}\tau _p}
\prod _{j=1}^{b_p}\int _{0}^1dw_j\ e^{2\pi\sqrt{-1}w_j\int _{\Sigma}\rho _j^{(p)}} \\ &=\cases 0,&\text{if $\int _{\Sigma}\rho _j^{(p)}\neq 0$
for any $j=1,\ldots ,b_p$}\\ \hat u_0^c(\Sigma )e^{2\pi\sqrt{-1}\int _{\Sigma}\tau _p},&\text{if $\int _{\Sigma}\rho _j^{(p)}=0$ for any
$j=1,\ldots ,b_p$}\endcases\endsplit\tag6.10$$ Using (6.2) the sum over the cohomology classes $c$ in (4.28) can be easily calculated

$$\multline\sum _{c\in H^{p+1}(M,\Bbb Z)}\hat u_0^c(\Sigma )e^{-\frac{1}{2}\|\delta _1(\hat u_0^c)\|^2}= \\ =\sum _{m_1\in\Bbb Z}\cdots
\sum _{m_{b_{p+1}}\in\Bbb Z} \prod _{j=1}^{b_{p+1}}(\hat f_j^{(p)}(\Sigma ))^{m_j}e^{-\frac{1}{2}\sum
_{i,j=1}^{b_{p+1}}h_{ij}^{(p+1)}m_{i}m_{j}}\sum _{y_1=1}^{l_1-1}\cdots\sum _{y_r=1}^{l_r-1}\prod _{k=1}^{r}(\hat t_k^{(p)}(\Sigma
))^{y_k}\endmultline\tag6.11$$ where

$$\sum _{y_1=1}^{l_1-1}\cdots\sum _{y_r=1}^{l_r-1}\prod _{k=1}^{r}(\hat t_k^{(p)}(\Sigma
))^{y_k}=\cases 0 &\text{if $\hat t_k^{(p)}(\Sigma )\neq 1$ for any $k=1,\ldots ,r$}\\ ord(TorH^{p+1}(M;\Bbb Z)) &\text{if $\hat
t_k^{(p)}(\Sigma )=1$ for any $k=1,\ldots ,r$}\endcases\tag6.12$$ In order to obtain a non-vanishing VEV of the Wilson operator, (6.1) implies
that the free part of $[\Sigma ]$ vanishes in $H_p(M;\Bbb Z)$ and by (6.12) that $\hat t_k^{(p)}(\Sigma )=1$ for all $k$. Let us now assume
that $[\Sigma ]\in TorH_{p}(M;\Bbb Z)$ then the condition

$$\hat t_k^{(p)}(\Sigma )=v_k|_{Z_p(M;\Bbb Z)}(\Sigma )=<[v_k],[\Sigma ]>=1\quad\forall k=1,\ldots ,r\tag6.13$$ implies that $[v_k]$ vanishes
on the torsion elements in $H_{p}(M;\Bbb Z)$ for all $k$. Repeating the arguments which lead to proposition 6.1 and using (2.3), it can be
shown that $[v_k]\in\ker\delta ^{\ast}$. However, according to (6.1) we have $t_k^{(p+1)}=\delta _2(\hat t_k^{(p)})=-\delta ^{\ast} [v_k]\neq
0$. So one has to draw the following conclusion:

\proclaim{Proposition 6.2} The VEV of the Wilson operator $W(\Sigma )$ vanishes, unless $[\Sigma ]=0$ in $H_{p}(M;\Bbb Z)$.\qed\endproclaim

Let us now turn to an explicit computation of $\Cal E^{(p)}(W(\Sigma ))$, where $\Sigma =\partial\Sigma ^{\prime}$ for $\Sigma ^{\prime}\in
C_{p+1}(M;\Bbb Z)$. This implies $\hat f_j^{(p)}(\partial \Sigma ^{\prime})=q(\int _{\Sigma ^{\prime}}\rho _{j}^{(p+1)})$ for all $j=1,\ldots
,b_{p+1}$. Concerning (4.28) it remains to perform an integration over $\tau\in imd_{p+1}^{\ast}$. The mapping $J_{\Sigma}\colon\tau
_p\rightarrow\int _{\Sigma}\tau _p$, which appears in the exponent of (6.10), can be regarded as a continuous linear functional on $\Omega
^{p}(M;\Bbb R)$, i.e. $J_{\Sigma}$ is a current of degree $(n-p)$.\par

We will briefly recall the main facts concerning the theory of currents [27]. Let us denote the space of $p$-currents by $\Cal C^p(M)$. Every
$\beta\in\Omega ^{p}(M;\Bbb R)$ gives rise to a $p$-current $\tilde\beta (\phi ):=\int _{M}\beta\wedge\phi$ for all $\phi\in\Omega
^{n-p}(M;\Bbb R)$. The differential of a $p$-current $T$ is the $(p+1)$-current $\tilde dT$, which is defined by $\tilde dT(\phi):=(1)^{p+1}
T(d\phi)$ for $\phi\in\Omega ^{n-p-1}(M;\Bbb R)$. The Hodge star operator naturally extends to an operator $\star\Cal C^{p}(M)\rightarrow\Cal
C^{n-p}(M)$ by $\star T(\phi ):=(-1)^{p(n-p)}T(\star\phi )$ for all $\phi\in\Omega ^{p}(M;\Bbb R)$. Subsequently one can introduce the
co-differential $\tilde d^{\ast}:=(-1)^{n(p+1)+1}\star \tilde d\star\colon\Cal C^p(M)\rightarrow\Cal C^{p-1}(M)$ on currents. Let
$<T,\phi>:=T(\star\phi )$ denote the scalar product of a $p$-current $T$ with a differential form $\phi\in\Omega ^{p}(M;\Bbb R)$. The Hodge
decomposition theorem extends to currents [27], so that every current $T$ can be uniquely written as

$$T=\tilde d_{p-1}\tilde d_{p}^{\ast}\tilde G_pT+\tilde d_{p+1}^{\ast}\tilde d_{p}\tilde G_pT+\widetilde{HT},\tag6.14$$ where the
Green\rq s operator $\tilde G_p$ on currents is defined by $<\tilde G_pT,\phi >=<T,G_p\phi >$ and $HT\in Harm^p(M)$ fulfills $<\widetilde
{HT},\rho >=<T,\rho>$ for all $\rho\in Harm^p(M)$.\par

Let us now introduce the dual current $j_{\Sigma}:=\star J_{\Sigma}\in\Cal C^p(M)$. Since $\Sigma =\partial\Sigma ^{\prime}$ one gets
$j_{\Sigma}=\tilde d_{p+1}^{\ast}j_{\Sigma ^{\prime}}$, so that

$$J_{\Sigma}(\tau _p)=\int _{\Sigma}\tau _p=<j_{\Sigma},\tau _p>=<j_{\Sigma ^{\prime}},d_p\tau _p>.\tag6.15$$ Performing formally the
Gaussian functional integration over $imd_{p+1}^{\ast}$ and using (6.10) and (6.11) one obtains the VEV of the Wilson operator

$$\Cal E^{(p)}(W(\Sigma )) =\frac{\Theta _{b_{p+1}} \left(\hat K^{(p+1)}(\Sigma^{\prime})|-\frac{h^{(p+1)}}{2\pi\sqrt{-1}}\right)}
{\Theta _{b_{p+1}} \left(0|-\frac{h^{(p+1)}}{2\pi\sqrt{-1}}\right)}\cdot e^{-\frac{(2\pi )^2}{2}<j_{\Sigma ^{\prime}},\tilde d_{p}\tilde
G_{p}\tilde d_{p+1}^{\ast}j_{\Sigma ^{\prime}}>},\tag6.16$$ where $\hat K_j^{(p+1)}(\Sigma^{\prime}):=\int _{\Sigma^{\prime}}\rho _j^{(p+1)}$
with $j=1,\ldots b_{p+1}$ is regarded as $b_{p+1}$-dimensional vector. Let us remark that one has to regularize $<j_{\Sigma ^{\prime}},\tilde
d_{p}\tilde G_{p}\tilde d_{p+1}^{\ast}j_{\Sigma ^{\prime}}>$, since the product of two currents is ill-defined in general. Making precise sense
of this inner product lies beyond the scope of this paper. However, in the example below we will give an explicit and finite expression.\par

Before closing this section, we want to notice that the value of $\Cal E^{(p)}(W(\Sigma ))$ does not depend on the choice for $\Sigma
^{\prime}$. In fact, suppose that there exists a second $(p+1)$-chain $\Sigma ^{\prime\prime}$, with $\Sigma =\partial\Sigma ^{\prime\prime}$,
then $\tilde K^{(p+1)}(\Sigma ^{\prime\prime} )-\tilde K^{(p+1)}(\Sigma ^{\prime})\in\Bbb Z$. The properties of $\Theta _{b_{p+1}}$ and the
fact that $\tilde d_{p+1}^{\ast}j_{\Sigma ^{\prime\prime}}=\tilde d_{p+1}^{\ast}j_{\Sigma ^{\prime}}$ proves this statement.
\bigskip
\subheading {The Wilson operator in codimension 1} As an example we will consider the VEV of $W(\Sigma )$ in codimension 1. We realize the
0-current $J_{\Sigma ^{\prime}}$ by the characteristic function $\nu _{\Sigma ^{\prime}}$ for $\Sigma ^{\prime}\subset M$. Hence $j_{\Sigma
^{\prime}}(\phi )=\int _M vol_M\phi\nu _{\Sigma ^{\prime}}$ for $\phi\in C^{\infty}(M)$. The main point is that $j_{\Sigma ^{\prime}}$ is
square summable $n$-current [27], hence its norm $\|j_{\Sigma ^{\prime}}\|^{2}=\int _{\Sigma ^{\prime}}vol_M=:Vol(\Sigma ^{\prime})$ is finite.
With respect to the normalized basis $\varrho ^{\prime (n)}=\frac{vol_M}{\sqrt{Vol(M)}}$ of $Harm ^n(M)$, one finds $Hj_{\Sigma
^{\prime}}=\frac{Vol(\Sigma ^{\prime})}{Vol(M)}vol_M$. Using the Hodge decomposition (6.14), the exponent in (6.16) becomes

$$<j_{\Sigma ^{\prime}},\tilde d_{n-1}\tilde G_{n-1}\tilde d_{n}^{\ast}j_{\Sigma ^{\prime}}>=
\|j_{\Sigma ^{\prime}}\|^{2}-<\widetilde{Hj_{\Sigma ^{\prime}}},Hj_{\Sigma ^{\prime}}>= Vol(\Sigma ^{\prime})\left( 1-\frac{Vol(\Sigma
^{\prime})}{Vol(M)}\right).\tag6.17$$ Applying the Poisson formula $\sum _{m=-\infty}^{\infty}G(m)=\sum _{k=-\infty}^{\infty}\int
_{-\infty}^{\infty}dyG(y)e^{-2\pi\sqrt{-1}ky}$, where $G(m)$ is an arbitrary function, one obtains

$$\Theta _1\left(\frac{Vol(\Sigma ^{\prime})}{Vol(M)}|-\frac{1}{2\pi\sqrt{-1}Vol(M)}\right)=
\Theta _1\left(1-\frac{Vol(\Sigma ^{\prime})}{Vol(M)}|-\frac{1}{2\pi\sqrt{-1}Vol(M)}\right).\tag6.18$$ Using (6.17) and (6.18) we finally get
from (6.16)

$$\Cal E^{(n-1)}(W(\Sigma ))=\frac{\Theta _1\left(1-\frac{Vol(\Sigma
^{\prime})}{Vol(M)}|-\frac{1}{2\pi\sqrt{-1}Vol(M)}\right)}{\Theta _1\left(0|-\frac{1}{2\pi\sqrt{-1}Vol(M)}\right)}\exp{\left( -\frac{(2\pi
)^2}{2}Vol(\Sigma ^{\prime})(1-\frac{Vol(\Sigma ^{\prime})}{Vol(M)})\right)},\tag6.19$$ showing that the VEV of $W(\Sigma )$ depends on the
volumes of $\Sigma ^{\prime}$ and $M$ only. Furthermore the result (6.19) is invariant under the replacement of $Vol(\Sigma ^{\prime})$ by
$Vol(M)-Vol(\Sigma ^{\prime})$. This symmetry can be traced back to the ambiguity in defining the boundary $\Sigma ^{\prime}$ of $\Sigma$,
since one could alternatively replace $\Sigma ^{\prime}$ by its complement $M\backslash\Sigma ^{\prime}$.
\bigskip\bigskip
{\bf Acknowledgments}
\bigskip I would like to express my gratitude to H. H\"{u}ffel for his encouragement and his comments.


\bigskip\bigskip

\Refs
\bigskip

\ref \no 1\by M. Kalb and P. Ramond\paper Classical direct interstring action\jour Phys. Rev.\vol D9 \yr 1974\pages 2273\endref

\ref \no 2\by E. Cremmer and J. Scherk\paper Spontaneous dynamical breaking of gauge symmetry in dual models\jour Nucl. Phys.\vol B72 \yr
1974\pages 117\endref

\ref \no 3\by D.S. Freed\paper Dirac charge quantization and generalized differential cohomology\jour In Surveys in differential geometry,
Surv. Differ. Geom.\vol VII\pages 129\finalinfo International Press, Somerville, MA, 2000\endref

\ref \no 4\by M.J. Hopkins and I.M. Singer\paper Quadratic functions in geometry, topology, and M-theory\jour J. Diff. Geom.\vol 70\yr 2005
\pages 329\endref

\ref \no 5\by J. Cheeger and J. Simons\paper Differential characters and geometric invariants; Geometry and Topology (College Park
Md.,1983/84)\jour Lecture Notes in Mathematics\vol 1167 \yr 1985 \pages 50\endref

\ref \no 6\by D.S. Freed, G.W. Moore and G. Segal\paper Heisenberg groups and noncommutative fluxes\jour Annals Phys.\vol 322\yr 2007\pages
236\endref

\ref \no 7\by D.S. Freed, G.W. Moore and G. Segal\paper The uncertainty of fluxes\jour Commun. Math. Phys.\vol 271\yr 2007\pages 247\endref

\ref \no 8\by D. Belov and G.W. Moore\paper Conformal blocks for $AdS_5$ singletons\jour arXiv: hept-th/0412167\endref

\ref \no 9\by E. Diaconescu, G.W. Moore and D.S. Freed\paper The M-theory 3-form and $E_8$ gauge theory\jour arXiv: hept-th/0312069 \endref

\ref \no 10\by G.W. Moore and E. Witten\paper Self-duality, Ramond-Ramond fields, and $K$-theory\jour JHEP\vol 0005\yr 2000\pages 032 \endref

\ref \no 11\by D. Freed and M. Hopkins\paper On Ramond-Ramond fields and $K$-theory\jour JHEP\vol 0005\yr 2000\pages 044\endref

\ref \no 12\by P.K. Townsend \paper Covariant quantization of antisymmetric tensor gauge fields\jour Phys. Lett.\vol 88B\yr 1979 \pages
97\endref

\ref \no 13\by W. Siegel\paper Hidden ghosts\jour Phys. Lett.\vol 93B\yr 1980 \pages 170\endref

\ref \no 14\by T. Kimura\paper Antisymmetric tensor gauge fields in general covariant gauge\jour Prog. Theo. Phys.\vol 64 \yr 1980\pages
357\endref

\ref \no 15\by Yu. N. Obuhkov\paper The geometrical approach to antisysmmetric tensor field theory\jour Phys. Lett.\vol 109B\yr 1982 \pages
195\endref

\ref \no 16\by A.S. Schwarz\book Quantum Field Theory and Topology\publ Springer\publaddr Berlin, Heidelberg, New York\yr 1993\endref

\ref \no 17\by A.S. Schwarz\paper The partition function of a degenerate functional\jour Commun. Math. Phys.\vol 67 \yr 1979\pages 1\endref

\ref \no 18\by A.S. Schwarz and Y.S. Tyupkin\paper Quantization of antisymmetric tensors and Ray-Singer torsion\jour Nucl. Phys.\vol B242 \yr
1984\pages 436\endref

\ref \no 19\by H. H{\"u}ffel and G. Kelnhofer\paper Generalized stochastic quantization of Yang-Mills theory\jour Ann. of Phys.\vol 270 \yr
1998\pages 231\endref

\ref \no 20\by H. H{\"u}ffel and G. Kelnhofer\paper Global path integral quantization of Yang-Mills theory\jour Phys. Lett.\vol B472 \yr
2000\pages 101\endref

\ref \no 21\by K. Gomi\paper Projective unitary representations of smooth Deligne cohomology groups\jour arXiv: math.RT/0510187 \endref

\ref \no 22\by R. Harvey, B. Lawson and J. Zweck\paper The de Rham-Federer theory of differential characters and character duality\jour Amer.
J. Math.\vol 125 \yr 2003 \pages 791\endref

\ref \no 23\by F.W. Warner\book Foundations of Differentiable Manifolds and Lie Groups\publ Springer\publaddr Berlin, Heidelberg, New York\yr
1983\endref

\ref \no 24\by G.E. Bredon\book Topology and Geometry\publ Springer\publaddr Berlin, Heidelberg, New York\yr 1992\endref

\ref \no 25\by J. Gegenberg and G. Kunstatter\paper The partition function for topological field theories\jour Annals Phys.\vol 231\yr 1994
\pages 270\endref

\ref \no 26\by D.B. Ray and I.M. Singer\paper R-Torsion and the Laplacian on Riemannian manifolds\jour Adv. Math.\vol 7 \yr 1979\pages
145\endref

\ref \no 27\by G. de Rham\book Differentiable manifolds, Forms, Currents, Harmonic Forms\publ Springer\publaddr Berlin, Heidelberg, New York\yr
1984\endref


\endRefs
\enddocument